\documentclass[fleqn,usenatbib,useAMS]{mnras}

\usepackage[dvipsnames]{xcolor}
\usepackage{tikz,hyperref}

\definecolor{lime}{HTML}{A6CE39}
\DeclareRobustCommand{\orcidicon}{
	\begin{tikzpicture}
	\draw[lime, fill=lime] (0,0) 
	circle [radius=0.13] 
	node[white] {{\fontfamily{qag}\selectfont \tiny ID}};
	\draw[white, fill=white] (-0.0625,0.095) 
	circle [radius=0.007];
	\end{tikzpicture}
	\hspace{-2mm}
}

\foreach \x in {A, ..., Z}{\expandafter\xdef\csname orcid\x\endcsname{\noexpand\href{https://orcid.org/\csname orcidauthor\x\endcsname}
			{\noexpand\orcidicon}}
}

\usepackage{newtxtext,newtxmath}

\usepackage{booktabs}

\usepackage[T1]{fontenc}
\usepackage{ae,aecompl}

\usepackage{graphicx}	
\usepackage{amsmath}	
\usepackage{multicol}
\usepackage{multirow}
\usepackage{graphicx}
\usepackage[dvipsnames]{xcolor}
\usepackage{booktabs}
\usepackage{times}
\input{epsf}

\title[3D physical structure and angular expansion of T\,Pyx]{3D physical structure and angular expansion of the remnant of the recurrent nova T\,Pyx}

\author[E.~Santamar\'{i}a et al.]{E.~Santamar\'{i}a\thanks{E-mail:\,e.santamaria@irya.unam.mx}$^{1\orcidA}$, J.~A.~Toal\'{a}$^{1\orcidB}$, M.~A.~Guerrero$^{2\orcidC}$, G.~Ramos-Larios$^{3\orcidD}$ and L.~Sabin$^{4\orcidE}$\\ 
$^{1}$Universidad Nacional Aut\'{o}noma de M\'{e}xico, Instituto de Radioastronom\'{i}a y Astrof\'{i}sica, 58090 Morelia, Michoac\'{a}n, Mexico\\
$^2$Instituto de Astrof\'{i}sica de Andaluc\'{i}a, IAA-CSIC, Glorieta de la Astronom\'{i}a S/N, Granada 18008, Spain \\
$^{3}$Instituto de Astronom\'{i}a y Meteorolog\'{i}a, CUCEI, Universidad de Guadalajara, Av. Vallarta 2602, Col. Arcos Vallarta, 44130 Guadalajara, Mexico\\
$^{4}$Universidad Nacional Aut\'{o}noma de M\'{e}xico, Instituto de Astronom\'{i}a, A.P. 877, 22800 Ensenada, B.C., Mexico
}

\date{\today}

\pubyear{2023}

\begin{document}
\label{firstpage}
\pagerange{\pageref{firstpage}--\pageref{lastpage}}
\maketitle

\begin{abstract}
\noindent We present the analysis of archival VLT MUSE and multi-epoch {\it HST} WFPC2 and WFC3/UVIS narrow-band observations of the remnant associated with the ejecta of the mid-nineteenth century outburst of the recurrent nova T\,Pyx.  
These data sets are used to investigate its true 3D physical structure and the nebular expansion patterns along the line of sight and on the plane of the sky.  
The VLT MUSE emission line maps and 3D visualisations based on position-position-velocity diagrams reveal T\,Pyx as a bipolar nebula, with a knotty toroidal structure at its waist best seen in H$\beta$ and two open bowl-shaped bipolar lobes (a diabolo) best revealed by the [O~{\sc iii}] emission lines. 
The comparison of multi-epoch \emph{HST} WFPC2 and WFC3/UVIS narrow-band images and VLT MUSE emission line maps of T\,Pyx reveals the angular expansion of the remnant through the proper motion of individual knots and nebular features. 
The angular expansion is confirmed to be homologous in the period from 1994.2 to 2007.4 before the recent 2011 outburst, but there is suggestive evidence that the inner knots have experienced a higher expansion rate since then.

\end{abstract}




\section{Introduction}
\label{sec:intro}

Among the eruptive and violent phenomena in the Universe, cataclysmic variables stars are worth contenders, where classical novae (CNe) eruptions are produced.  
Currently there are only 10 confirmed recurrent novae (RNe) in our Galaxy, systems that are known to experience multiple nova eruptions. 
These systems are composed by a relatively massive ($M_\mathrm{WD}\gtrsim1$ M$_\odot$) white dwarf (WD) accreting material from a Roche-lobe-filling, main-sequence companion \citep{Knigge2006}. 
When the pressure of the material accumulated at the surface of the WD reaches critical values, a thermonuclear explosive hydrogen-burning event takes place ejecting 10$^{-7}$ to 10$^{-3}$ M$_\odot$ of material with expansion velocities from several hundred to a few thousand km~s$^{-1}$ \citep[see][and references therein]{1989PASP..101....5S, 2008clno.book.....B}. 
RNe represent the only class of novae whereby the WD has the potential to increase its mass up to the critical value for a supernova event, making them the most promising candidates for Type Ia supernova progenitors.

Among these objects, T\,Pyx was the first recurrent nova ever discovered \citep{Leavitt1913}, and it has thus received significant attention \citep[e.g.,][and references therein]{Knigge2022}. 
Historical records indicate notable outbursts occurred in 1890, 1902, 1920, 1944 and 1966, with a delayed eruption that took place on 2011 April 14 \citep{2011CBET.2700....1W}. 
Excluding this last eruption, the outburst periodicity of T\,Pyx would be 19$\pm$5.3 yr \citep[][]{1987ApJ...314..653W}. 
\citet{2013ApJ...770L..33S} estimated a distance of 4.8$\pm$0.5 kpc based on the propagation of light echoes from the 2011 outburst through the nebular material associated with previous eruptions, while \citet{Izzo2023} estimated a distance of 3.55$\pm$0.77~kpc using an expansion parallax method. 
The distance determination of 2.6$^{+0.4}_{-0.2}$~kpc using {\it Gaia} data \citep{BJ2021} that will be adopted here is notably smaller.
T\,Pyx is considered to be a moderately fast nova, with $t_{2}=31$ d and $t_{3}=62$ d \citep{2010ApJS..187..275S}\footnote{$t_2$ and $t_3$ are the times it takes the nova outburst to decline 2 and 3 mag from the peak brightness, respectively \citep[see][]{Gaposchkin1957}.}. 
The binary period of T\,Pyx is 1.8~h \citep{Patterson1998} and the accretion rate has been estimated to be $M_\mathrm{acc}=10^{-7}$~M$_\odot$~yr$^{-1}$ \citep{Patterson2017}. 
The quiescent bolometric luminosity of T\,Pyx is one order of magnitude higher than anticipated, leading to the hypothesis that it will soon undergo a potentially final and rapid phase of evolution \citep{2000A&A...364L..75K}.

The nova remnant of T\,Pyx was initially described as a circular ring of 5~arcsec in radius based on H$\alpha$-filter images obtained with photographic plates \citep{1979Duerbeck}. Subsequently deeper CCD H$\alpha$+[N~{\sc ii}] images revealed a faint halo extending up to 10~arcsec \citep{Shara+1989}.  
This nebular emission was then interpreted as a shell with limb-brightened enhancement, but it was finally resolved by \emph{Hubble Space Telescope} (\emph{HST}) images obtained through the H$\alpha$ F656N and most notably the [N~{\sc ii}] F658N filters \citep[][]{Shara+1997}. 
Those images revealed a myriad of bright knots radially distributed, very similar to the morphology observed in other nova-like objects  \citep[e.g., GK\,Per and IPHASX J210204.7+471015 ,][]{2012ApJ...761...34L,2018ApJ...857...80G}. Moreover, multi-epoch \emph{HST} observations in the light of H$\alpha$ of the T\,Pyx 2011 after eruption show the ionising flash of radiation \citep{Shara+2015}.

\begin{figure*}
\begin{center} 
\includegraphics[angle=0,width=\linewidth]{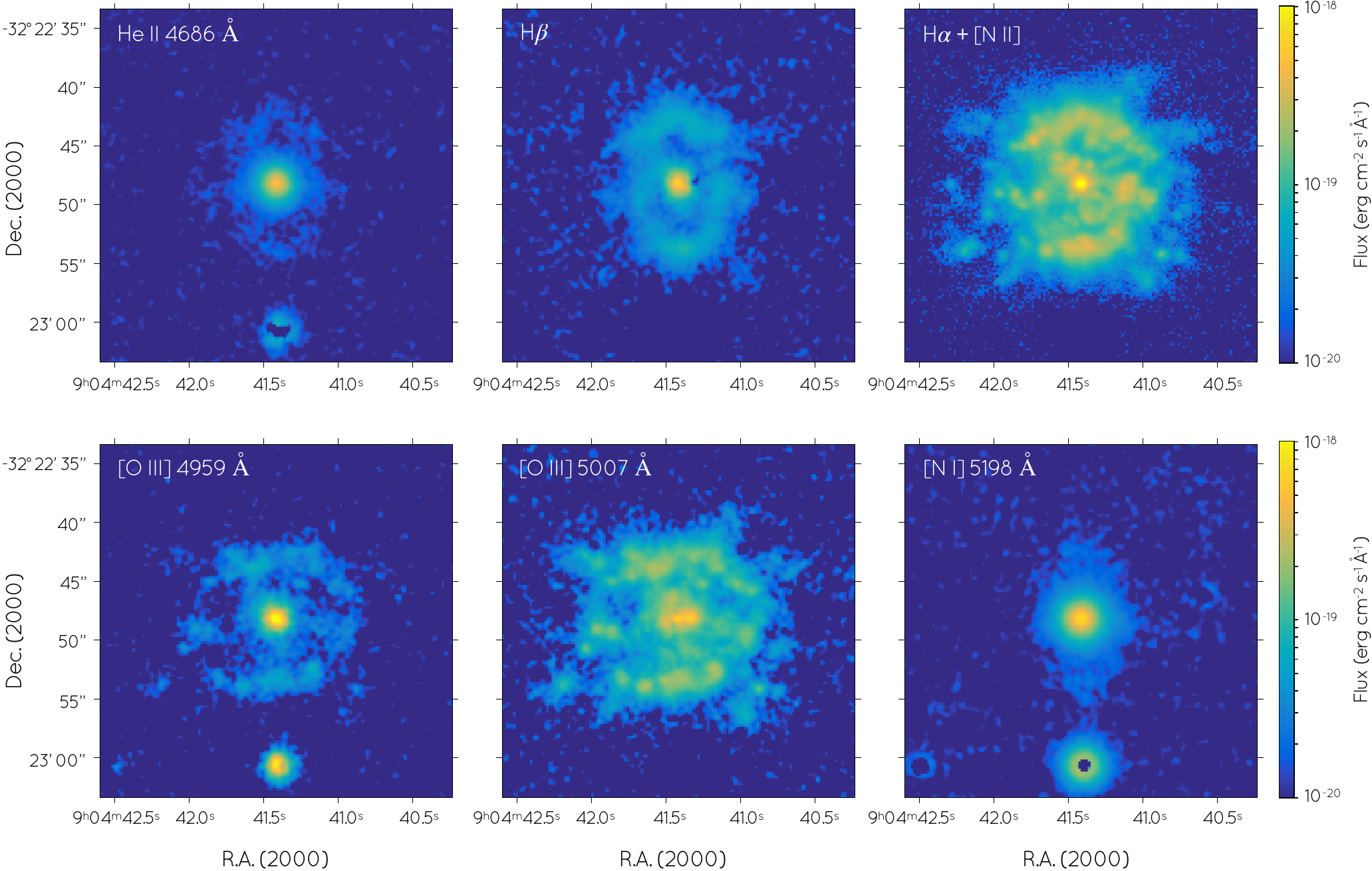}
\caption{T~Pyx VLT MUSE maps of the He~{\sc ii} $\lambda$4686, H$\beta$, and H$\alpha$+[N~{\sc ii}] $\lambda\lambda$6548,6584 (top row), and [O~{\sc iii}] $\lambda$4959, [O~{\sc iii}] $\lambda$5007 and [N~{\sc i}] $\lambda$5198+5200 (bottom row) emission lines. The bright source located $\sim$13~arcsec South of the central star of T\,Pyx in the He\,{\sc ii}, [O\,{\sc iii}] $\lambda$4959 and [N\,{\sc i}] maps is a background source. 
North is top, East to the left. 
The FoV is $\approx$30 arcsec in size. }
\label{fig:muse_images}
\end{center} 
\end{figure*}

The radial expansion velocity of the nova remnant of T\,Pyx, as determined from the [N~{\sc ii}] double-peaked line profile along a line of sight through its center, has been estimated to be 
$\simeq$500 km~s$^{-1}$ \citep{1998ApJ...498L..59O}, although the blend of the H$\alpha$ and [N~{\sc ii}] emission lines hampers a firm velocity determination.
Meanwhile, \citet{Schaefer2010} used multi-epoch \emph{HST} images obtained between 1994 and 2007 with the F658N narrow-band filter to assess the angular expansion of the nova remnant, reporting an expansion velocity in the plane of the sky for the material in the nova remnant between 500 and 715 km~s$^{-1}$ at their adopted distance of 3.5~kpc, i.e., tangential expansion velocities between 370 and 530 km~s$^{-1}$ at the \emph{Gaia} distance of 2.6 kpc. 
Those observations found no evidence of deceleration, suggesting the occurrence of a classical nova outburst in 1866 \citep{Schaefer2010}.

Thus far, the best way to unveil the true morpho-kinematics and energetics of nova remnants is the analysis of integral field spectroscopic (IFS) observations 
\citep[e.g.,][]{2009AJ....138.1090L,2009AJ....138.1541M,2023Celedon}. 
IFS data have the potential to determine the velocity field, true 3D shape, ionised mass and kinetic energy of nebular nova remnants \citep[e.g., as for the case of QU\,Vul,][]{Santamaria2022}. 
It also provides us with fine details of their clumpy structure that are associated with the early fragmentation of the ejecta, with great implications in our understanding of the formation and evolution of nova remnants. 
Particularly for T~Pyx, \citet{Izzo2023} has recently presented an analysis of IFS data to study the most recent 2011 ejection event using Very Large Telescope (VLT) Multi Unit Spectroscopic Explorer (MUSE) observations acquired with the MUSE Narrow-Field Mode (NFM) and VLT Adaptive Optics (AO).  
In conjunction with {\it HST} data, an upper limit to the total mass ejected during the 2011 event of 3$\times10^{-6}$~M$_\odot$ was estimated. 
Additional MUSE Wide-Field Mode (WFM) data suggest the presence of a ring-like structure from the mid-nineteenth century outburst with a radius of 5.6$\pm$0.7 arcsec  
tilted 63$^{\circ}$ from the plane of the sky and expanding at 471$^{+77}_{-72}$~km~s$^{-1}$ \citep{Izzo2023}.

In this paper, we present a reanalysis of the VLT MUSE WFM IFS data of T\,Pyx in conjunction with multi-epoch {\it HST} images to 
assess the expansion patterns along the line of sight and on the plane of the sky and hence determine its true 3D physical structure, kinematics and energetics. 
The paper is organised as follows. 
In Section~\ref{sec:obs} we present and describe the observations used in this paper. 
Section~\ref{sec:results} shows the results obtained from the analysis of the VLT MUSE and the multi-epoch analysis of {\it HST} data. A discussion of these results and the 3D structure unveiled by the observations is discussed in Section~\ref{sec:discussion}. 
A summary of our results is finally presented in Section~\ref{sec:summary}.

\section{Observations}
\label{sec:obs}

We retrieved calibrated VLT MUSE WFM observations of T\,Pyx from the ESO Archive Science Portal\footnote{\url{http://archive.eso.org/cms.html}}. The VLT MUSE data were taken on 2022 January 1 with a total exposure time of 5440~s (program ID 108.22A4, PI: L.\ Izzo). 
The observations cover the spectral range of 4600--9350~\AA~ with a spectral resolution $R\approx3000$. 
The data were processed with the pipeline of VLT MUSE \citep[see][for further details]{Weilbacher2020}.
The pixel size is 0.2$\times$0.2~arcsec$^2$ 
and the spatial resolution is $\simeq$1.0~arcsec, as derived from the FWHM of the central star of T\,Pyx.

The MUSE observations were complemented with available multi-epoch {\it HST} images retrieved from the Hubble Legacy Archive\footnote{\url{https://hla.stsci.edu/}}. 
The data used here correspond to narrow-band images in the [N~{\sc ii}] $\lambda$6584 emission line.  
Four epochs are currently available in the archive, three correspond to data obtained with the Wide Field Planetary Camera 2 (WFPC2) on 1994 Feb 26 (1994.16), 1995 Oct 7 (1995.75) and 2007 Jun 29 (2007.42) belonging to programs 5482, 6311 and 10834, respectively, and the fourth was taken with the Wide Field Camera 3 (WFC3) and the channel for ultraviolet and visible light (UVIS) on 2013 Sep 30 (2013.75) as part of program 13400. 
The WFPC2 F658N and WFC3/UVIS F658N filters were used, respectively.  
It is important to note, however, that the transmission curves of the WFPC2 F658N ($\lambda_\mathrm{c}$=6591.7 \AA, FWHM=39.2 \AA) and WFC3/UVIS F658N ($\lambda_\mathrm{c}$=6586.2 \AA, FWHM=27.5 \AA) filters are notably different.  

Details of the \emph{HST} observations are presented in Table~\ref{tab:images}. 

\begin{table}
\caption{Log of the {\it HST} observations used in this work.}
\label{images}
\centering 
\label{tab:images}
\begin{tabular}{clccc} 
\hline
Date &  \multicolumn{1}{c}{Instrument} & Filter  &  Exposure Time  & Pixel Scale    \\
     &             &         &  (s)  & (arcsec pix$^{-1}$)   \\
\hline
1994.16 &  WFPC2 & F658N & 2400 & 0.05 \\
1995.75 &  WFPC2 & F658N & 7400 & 0.05 \\
2007.42 &  WFPC2 & F658N & 1900 & 0.05 \\
2013.75 &  WFC3/UVIS  & F658N & 1260 & 0.05 \\
\hline
\end{tabular}
\end{table}

\section{Analysis}
\label{sec:results}

The IFS data cube and multi-epoch images of T\,Pyx provide us with the opportunity to investigate its kinematics (along the line of sight) and its tangential velocity on the plane of the sky (by the analysis of its angular expansion) 
to obtain a view of the 3D physical structure of its nova remnant without geometric assumptions. This can provide accurate information of the ejecta to assess its expansion and get insights on the physics of the recurrent outbursts of T\,Pyx. 

\subsection{IFS data}

The VLT MUSE observations of T\,Pyx detect the He~{\sc ii} $\lambda$4686, H$\beta$, H$\alpha$, [N~{\sc ii}] $\lambda\lambda$6548,6584, [O~{\sc iii}] $\lambda\lambda$4959,5007 and [N~{\sc i}] $\lambda\lambda$5198,5200 emission lines (Fig. \ref{fig:muse_images}).
The [N~{\sc i}] doublet is spectroscopically unresolved, whereas the large expansion velocity of T\,Pyx blends the H$\alpha$ and [N~{\sc ii}] $\lambda\lambda$6548,6584 emission lines.

\begin{figure}
\begin{center}
\includegraphics[width=1.0\linewidth]{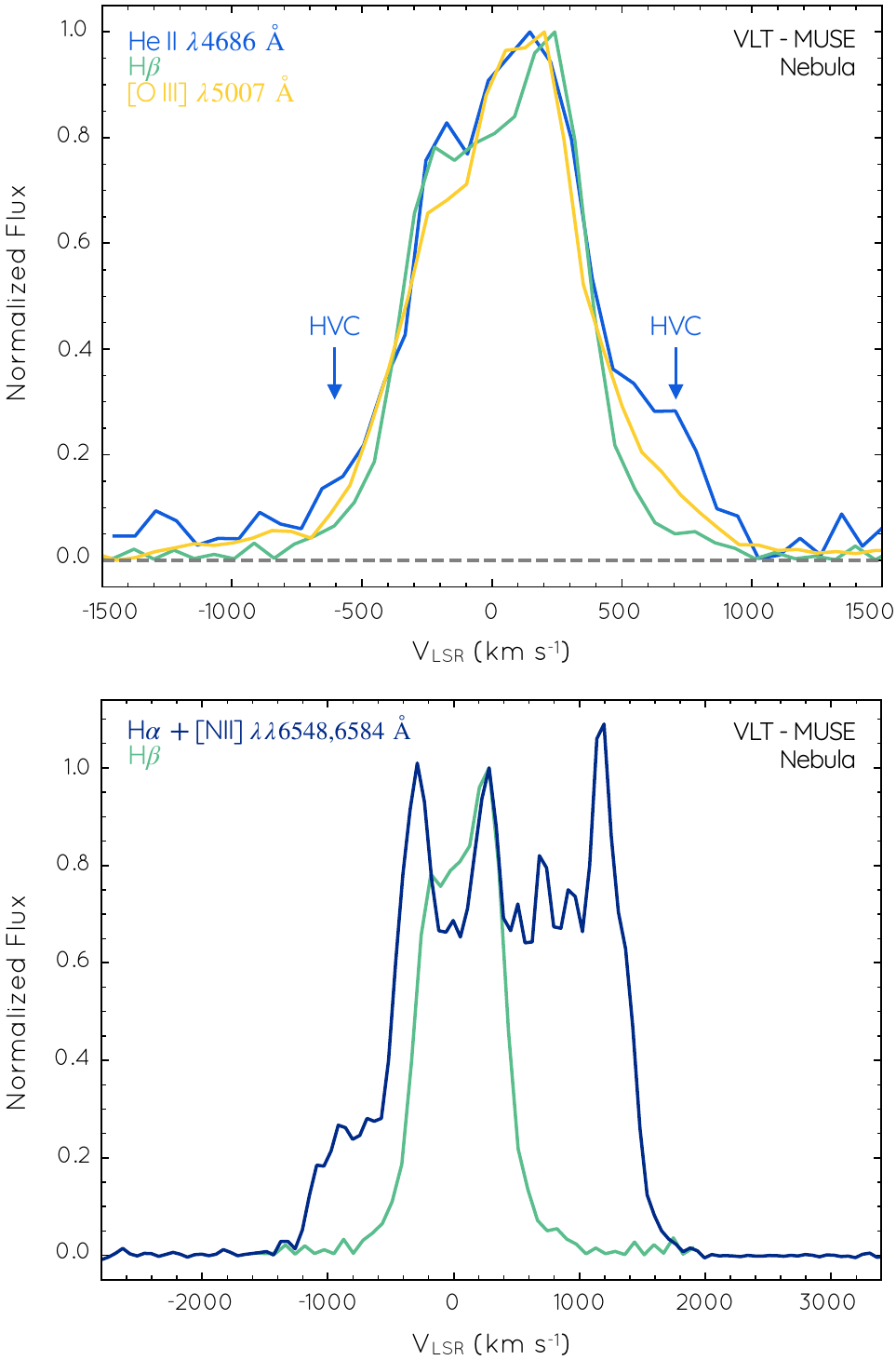}
\caption{
T\,Pyx normalised emission line profiles of bright lines in the MUSE data cube. 
(top) 
Profiles of the He~{\sc ii} $\lambda$4686, H$\beta$ and [O\,{\sc iii}] $\lambda$5007 emission lines. 
The line profile of [O\,{\sc iii}] and very particularly that of He~{\sc ii} show high-velocity components (HVCs) marked by blue arrows.  
The grey-dashed line shows the zero intensity level to emphasise the presence of these HVCs. 
(bottom) 
Profiles of the H$\alpha$ and [N\,{\sc ii}] $\lambda\lambda$6548,6584 emission lines. 
These are blended, given the large expansion velocity of T\,Pyx.  
The H$\beta$ profile is overlaid to indicate the possible contribution from the H$\alpha$.  
}
\label{fig:muse_lines}
\end{center}
\end{figure}

The H$\alpha$+[N~{\sc ii}] and [O~{\sc iii}] $\lambda$5007 emission lines result in the highest signal-to-noise (S/N) images as illustrated in Fig.~\ref{fig:muse_images}. 
These show an apparent ring-like structure, composed of dense and compact knots and filaments, with an averaged angular radius of about $\approx$5.6~arcsec and a width of $\sim$1~arcsec. 
Emission is detected up to radial distances $\approx$12 arcsec in these H$\alpha$+[N~{\sc ii}] and [O~{\sc iii}] images, disclosing extended emission not detected in {\it HST} images.
To a lesser extent, the [O\,{\sc iii}] $\lambda$4959 emission map is similar to that of [O\,{\sc iii}] $\lambda$5007, but with lower S/N ratio, as it could have been expected. 
The H$\beta$ emission line map presents a less complex morphology, with an apparent ellipsoidal structure oriented along position angle (PA) $\approx$+5$^\circ$ (East from the North direction) with semi-major and semi-minor axes of 5.6 and 3.5~arcsec, respectively. 
This morphology is rather consistent with that displayed by the lower S/N ratio emission maps of He\,{\sc ii} $\lambda$4686 and [N\,{\sc i}] $\lambda\lambda$5198,5200. 

The normalised spectral profiles of the nebular\footnote{
An irregular polygonal region was used to extract the nebular line profiles that included the entire nebular emission while excising the emission from T\,Pyx central star.} He~{\sc ii} $\lambda$4686, H$\beta$ and [O~{\sc iii}] $\lambda$5007 emission lines are presented in the top-panel of Fig.~\ref{fig:muse_lines}. 
These spectral profiles have FWHM of $\simeq$1000 km~s$^{-1}$, thus implying that the bulk of material is expanding at radial velocities up to
\begin{equation}
    V_{\rm exp} \approx \frac{\mathrm{FWHM}}{2\sqrt{\ln{2}}} \approx 600~{\rm km~s}^{-1}.
\end{equation}
The full width at zero intensities (FWZI) of these profiles actually extends up to $\simeq$1400 km~s$^{-1}$, with high-velocity components (HVCs), as marked in the top-panel of Fig.~\ref{fig:muse_lines}, that reveal that some material expands at higher velocities, up to $\pm$700~km~s$^{-1}$. 
These HVCs are detected most notably redwards of the He~{\sc ii} emission line, but also bluewards.  
They seem also present in the [O~{\sc iii}] emission line, but the emission of the HVCs is diluted by the much brighter emission in this line of the lower velocity material.

\begin{figure*}
\begin{center} 
\includegraphics[angle=0,width=.9\linewidth]{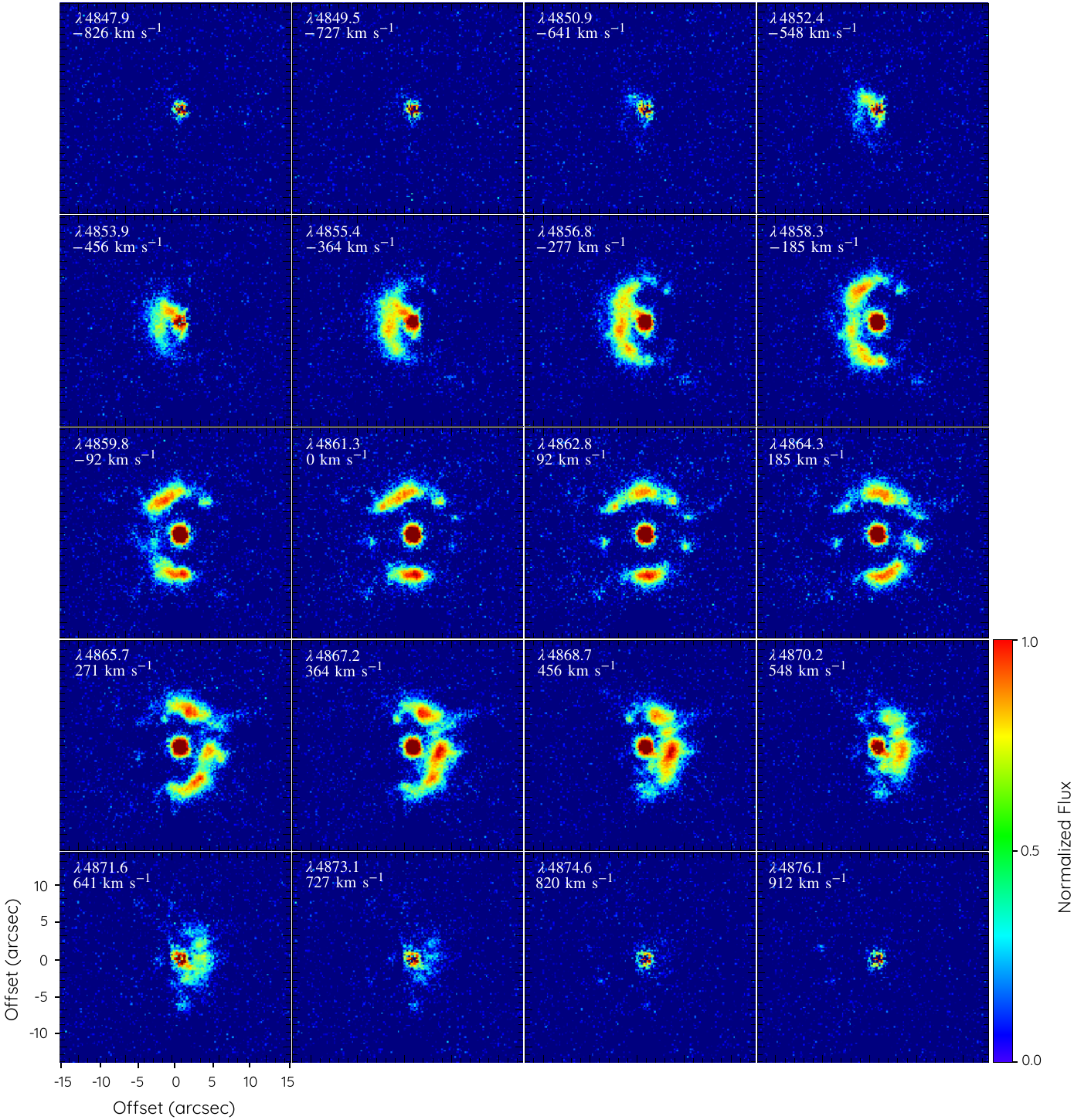}
\caption{
VLT MUSE H$\beta$ velocity channel maps of T\,Pyx. 
The wavelength and velocity of each channel map is indicated on each panel.  
The theoretical wavelength of the H$\beta$ emission line adopted here is 4861.332 \AA. 
Velocities are refered to the Local Standard of Rest (LSR). 
North is up, East to the left. 
}
\label{fig:channel3}
\end{center} 
\end{figure*}

\begin{figure*}
\begin{center} 
\includegraphics[angle=0,width=\linewidth]{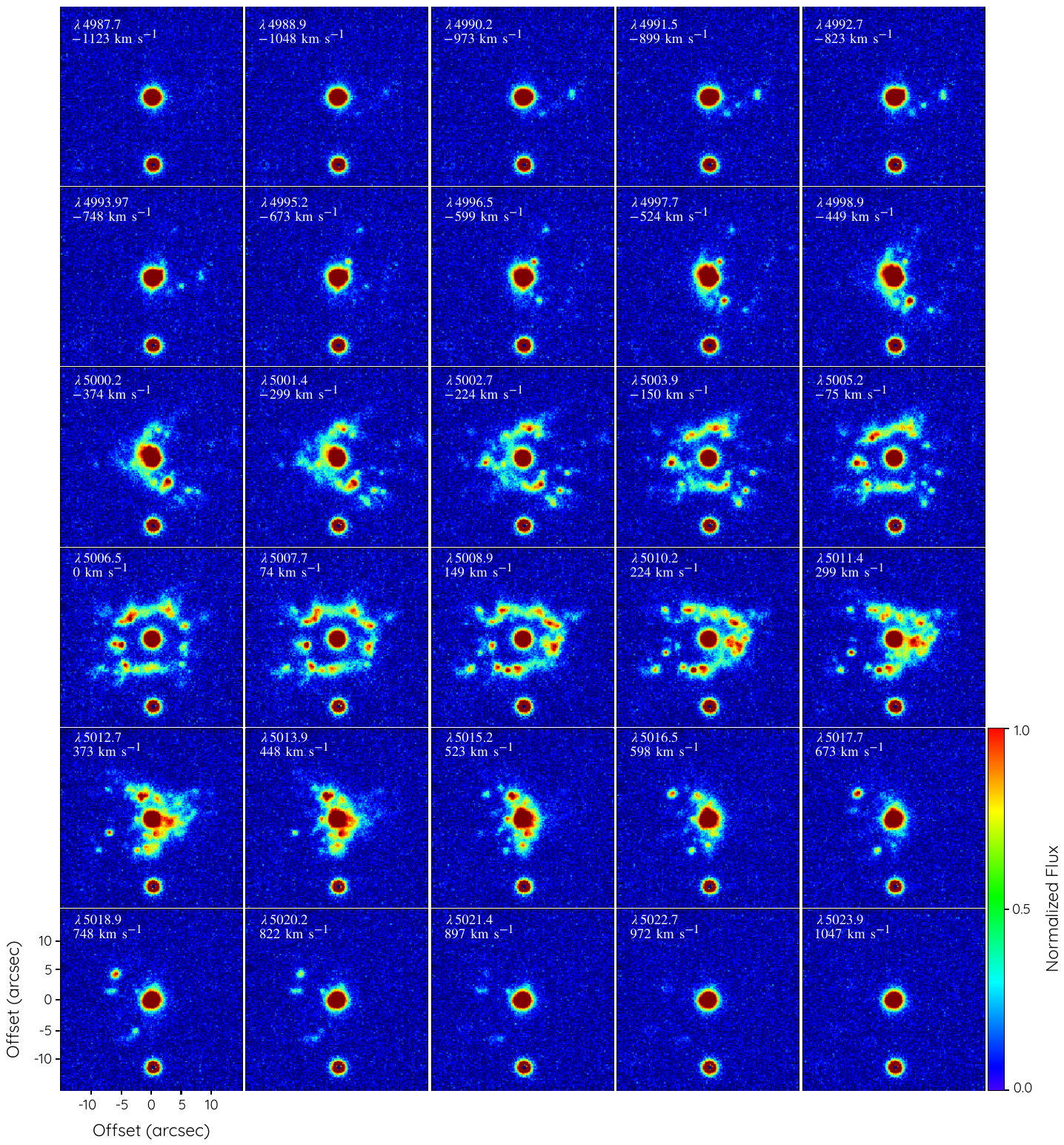}
\caption{
As in Fig.~\ref{fig:channel3}, VLT MUSE [O~{\sc iii}] $\lambda$5007 velocity channel maps of T\,~Pyx.  
The theoretical wavelength of the [O~{\sc iii}] $\lambda$5007 emission line adopted here is 5006.86~\AA. 
Note the background source located $\sim$13~arcsec South of T\,Pyx.}

\label{fig:channel2}
\end{center} 
\end{figure*}

\begin{figure*}
\begin{center} 
\includegraphics[angle=0,width=\linewidth]{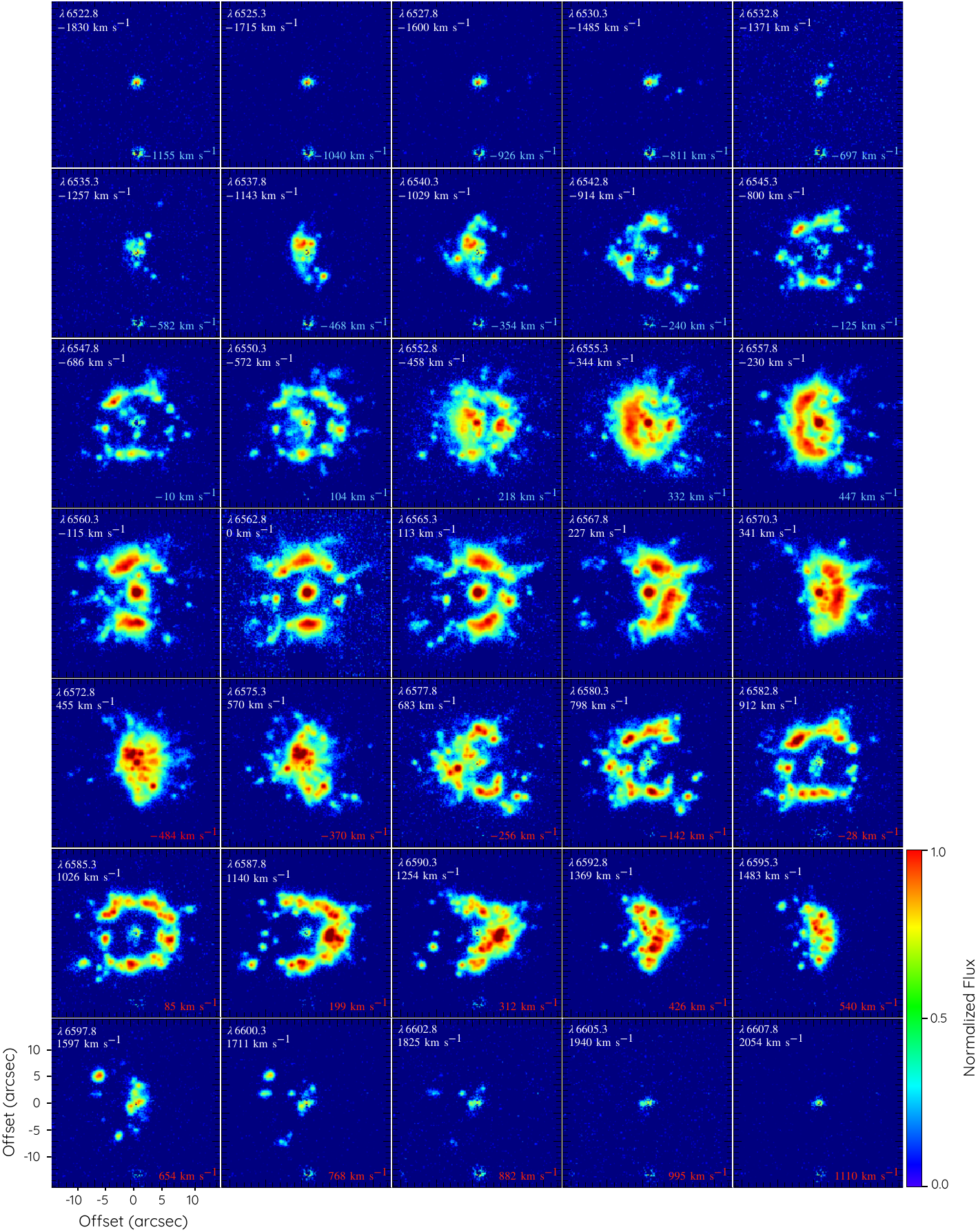}
\caption{
As in Figs.~\ref{fig:channel3} and \ref{fig:channel2}, VLT MUSE H$\alpha$ and [N\,{\sc ii}] $\lambda\lambda$6548,6584 velocity channel maps of T\,Pyx. 
The theoretical wavelength of the H$\alpha$ and [N~{\sc ii}] $\lambda\lambda$6548,6584 emission lines adopted here are 6562.849 \AA, 6548.05 \AA\ and 6583.45~\AA, respectively. 
The velocity of the H$\alpha$, [N~{\sc ii}] $\lambda$6548 and [N~{\sc ii}] $\lambda$6584 channel maps are labelled in white, cyan and red, respectively. 
}
\label{fig:channel1}
\end{center} 
\end{figure*}

The bottom panel of Fig.~\ref{fig:muse_lines} shows the normalised nebular spectral profile of the H$\alpha$ line. 
Given the large expansion velocity of this nova remnant, the H$\alpha$ line is blended with (at least\footnote{The possible presence of the He~{\sc ii} $\lambda$6560 emission line cannot be inferred as it would be overwhelmed by the much brighter H$\alpha$ and [N~{\sc ii}] emission lines.}) the [N~{\sc ii}] $\lambda\lambda$6548,6584 emission lines. 
The notorious blend between the H$\alpha$ and [N~{\sc ii}] emission lines is highlighted by the H$\beta$ emission line profile normalised to that of the peak of the H$\alpha$ profile overplotted on this figure, as the nebular H$\alpha$ and H$\beta$ line profiles can be expected to be consistent to each other. 
The subtraction of the normalised H$\beta$ profile allows recovering the possible ``clean'' profiles of the [N~{\sc ii}] emission lines (see Fig.~\ref{fig:subs} of Appendix \ref{app:over}), but the peak ratio of the emission profiles of the [N~{\sc ii}] $\lambda$6548 and [N~{\sc ii}] $\lambda$6584 do not comply the theoretical one-to-three ratio.
This situation hampers drawing firm conclusions on the overall kinematics of T\,Pyx using either the H$\alpha$ or the [N~{\sc ii}] emission lines \citep[see also the case presented in][]{Tappert2023}.

To peer into the morpho-kinematics of the nova remnant of T\,Pyx, we produced the channel maps of the H$\beta$, [O\,{\sc iii}] $\lambda$5007 and H$\alpha$+[N\,~{\sc ii}] emission lines shown in Figs.~\ref{fig:channel3}, \ref{fig:channel2} and \ref{fig:channel1}, respectively. 
The zero velocity channel map of each figure is referred to the Local Standard of Rest (LSR).  
Theoretical wavelengths of the H$\beta$ ($\lambda$4861.332~\AA), [O\,{\sc iii}] $\lambda$5007($\lambda$=5006.86~\AA), and H$\alpha$ ($\lambda$=6562.849~\AA), [N~{\sc ii}] $\lambda$6548 ($\lambda$=6548.05~\AA) and [N~{\sc ii}] $\lambda$6584 ($\lambda$=6583.45~\AA) emission lines have been adopted in Figs.~\ref{fig:channel3}, \ref{fig:channel2} and \ref{fig:channel1}, respectively.
The channel maps of the H$\beta$ and [O\,{\sc iii}] emission lines presented in Fig.~\ref{fig:channel3} and Fig.~\ref{fig:channel2} can be considered representative of the true morpho-kinematics of the nova remnant around T\,Pyx, given that they are not blended with any other emission line unlike the H$\alpha$+[N\,~{\sc ii}] emission lines.

The simplest case is that of the H$\beta$ emission line.  
It traces the brightest ellipsoidal structure in the nova remnant around T\,Pyx. 
The velocity channels maps of the H$\beta$ line exhibit very faint emission spanning from $-$727~km~s$^{-1}$ up to $+$912~km~s$^{-1}$, with emission blue-shifted towards the East of the central star and red-shifted towards its West (see Fig.~\ref{fig:channel3}). 
The emission seems consistent with a clumpy (or disrupted) tilted toroidal structure, with East blue-shifted emission peaking at $\approx-$277~km~s$^{-1}$ and West red-shifted emission peaking at $\approx+$460~km~s$^{-1}$, thus implying a systemic velocity at channel $\approx+$90~km~s$^{-1}$.  
Accordingly the radial expansion velocity would be $\simeq$370~km~s$^{-1}$. 
The semi-major and semi-minor axes of $5.6\pm0.4$ and $3.5\pm0.4$ arcsec of this toroidal structure imply an inclination of its symmetry axis with the line of sight $51^\circ\pm9^\circ$.  
Consequently, the true expansion velocity would be $470^{+80}_{-40}$ km~s$^{-1}$.

The channel maps of the [O\,{\sc iii}] $\lambda$5007 line in Fig.~\ref{fig:channel2} exhibit emission from a larger velocity range, from $-1070$ to $+1000$ km~s$^{-1}$. 
The channel maps in the central velocity range from $\approx-$300 to $\approx+$450 km~s$^{-1}$ are consistent with the toroidal structure disclosed by the H$\beta$ emission line in Fig.~\ref{fig:channel3}, but additional clumps beyond this structure are present suggesting more extended emission.  
Very noticeable these clumps show red- and blue-shifted emission both towards the East and West of the central star. 
Moreover a series of high-velocity clumps can be detected, with the most blue-shifted ones this time to the West of the central star and the most red-shifted ones to its East. 
These spatio-kinematic properties, particularly the reversal of the velocity sign with respect to the tilted toroidal structure, suggest clumps moving orthogonal to it. 
The channel map at $+$74 km~s$^{-1}$ can be considered representative of the systemic velocity, in agreement with that of $+$92 km~s$^{-1}$ derived from the H$\beta$ line.

Finally the channel maps of the blended H$\alpha$+[N\,~{\sc ii}] emission lines are shown in Fig.~\ref{fig:channel1}. 
Here the approaching material is located eastwards of the central star, while the receding material is located towards the West. 
The complete range in velocity of these blended emission lines goes from $-$1500 km~s$^{-1}$ in H$\alpha$ (although emission from the adjacent [N~{\sc ii}] $\lambda$6548 emission line at $-$735 km~s$^{-1}$ cannot be discarded) up to +1800~km~s$^{-1}$ in the [N~{\sc ii}] $\lambda$6584 emission line. 
The channel maps of the rest velocity of the H$\alpha$ and [N~{\sc ii}] emission lines ($v=0$ km~s$^{-1}$) disclose a clumpy morphology. 
As for the channel maps of the [O\,{\sc iii}] $\lambda$5007 emission line, the most blue-shifted clumps are found West of the central star and the most red-shifted clumps towards its East.

\begin{figure*}
\begin{center} 
\includegraphics[angle=0,width=0.85\linewidth]{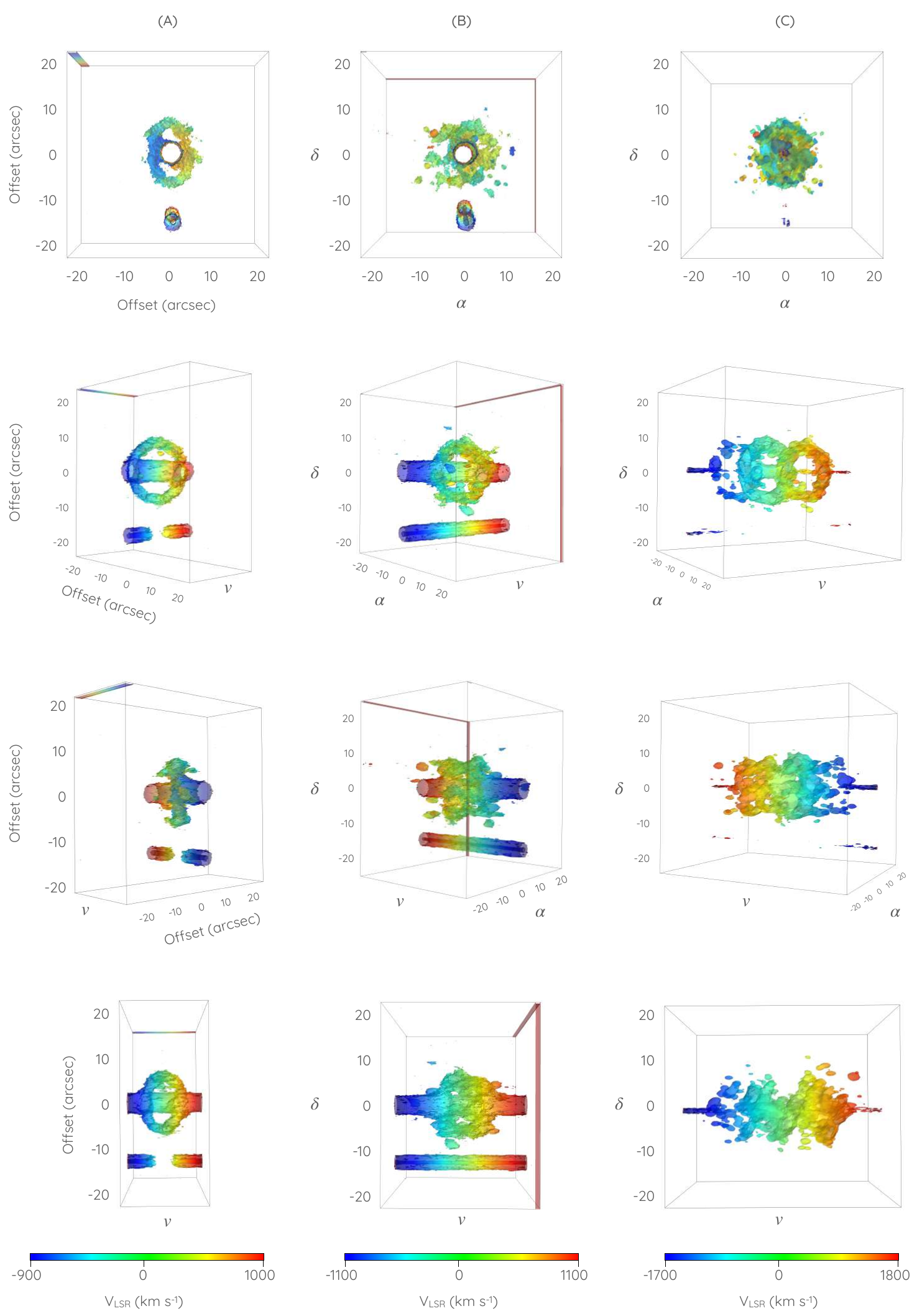}
\caption{
PPV diagrams of the H$\beta$ (column A), [O\,{\sc iii}] $\lambda5007$ \AA\, (column B) and H$\alpha+$[N\,{\sc ii}] $\lambda\lambda 6548,6584$ \AA\, (column C) emission lines of T Pyx extracted from the VLT MUSE datacube. 
The colour code represents velocity in the LSR. 
The top row shows the projection along the observer's point of view (i.e., the direct image), whereas the second and third rows show the projection from the plane of the sky at 45$^{\circ}$ to the West and East, respectively. 
Finally the fourth row shows a projection perpendicular to the plane of the sky to show the radial velocity (or $\lambda$). 
An animated figure of these panels is available in the online version.
Note that continuum dominated stellar sources, such as the central star of T\,Pyx and the background star towards its south (see Fig.~\ref{fig:channel2}), show in all panels as ``long, narrow cylinders''.  
}
\label{fig:3D}
\end{center} 
\end{figure*}

\begin{figure*}
\includegraphics[angle=0,width=0.9\linewidth]{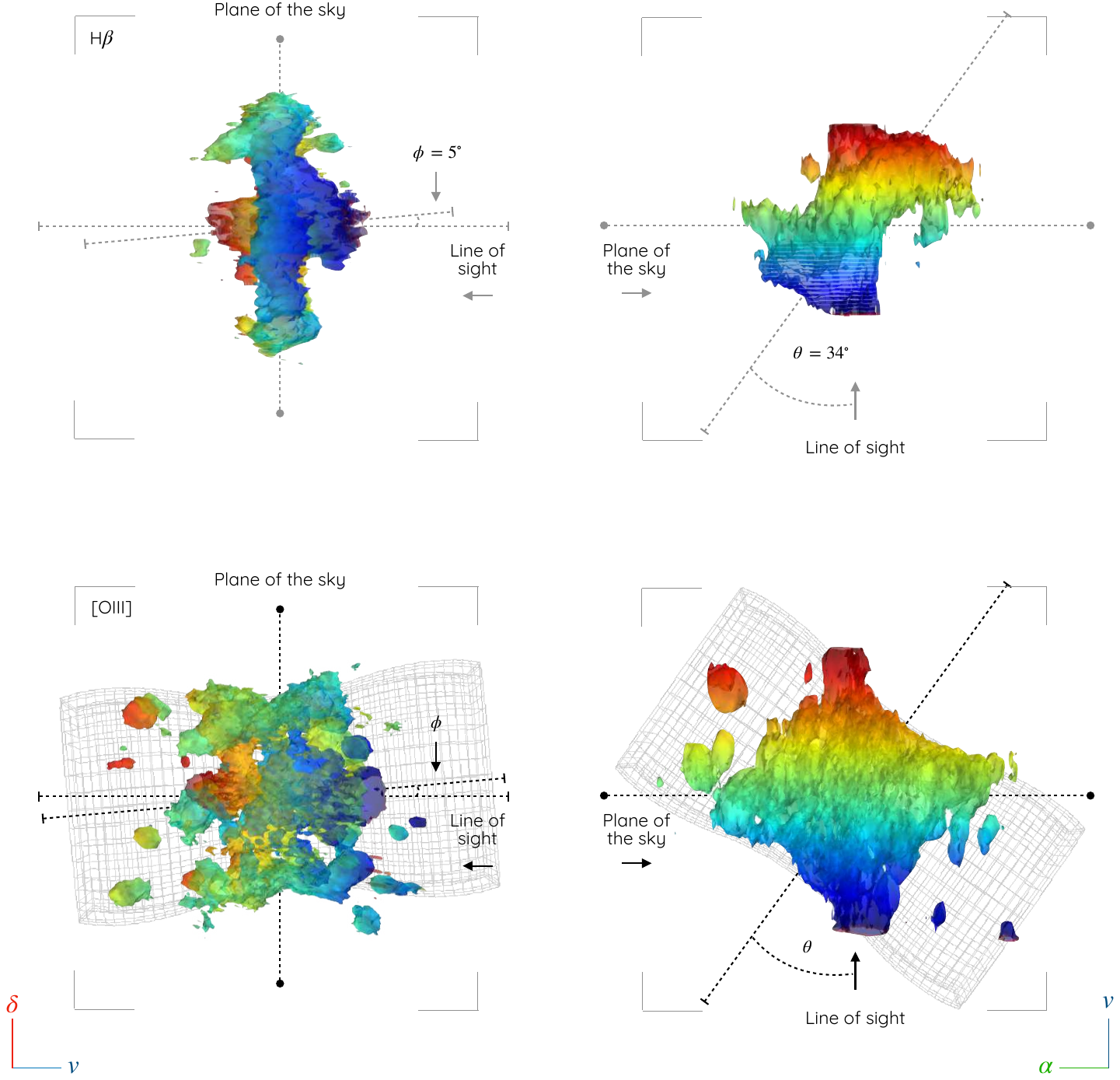}
\caption{
3D visualisation of the H$\beta$ (left column) and [O\,{\sc iii}] $\lambda5007$ \AA\, (right column) emission lines. 
A mesh pattern has been added to highlight the bipolar 3D structure of the ejecta. 
The colors represent the velocity field presented in Fig. \ref{fig:3D}.
}
\label{fig:inc}
\end{figure*}

To further examine the morpho-kinematics of this nova remnant, we show in Fig.~\ref{fig:3D} position-position-velocity (PPV) diagrams of the H$\beta$, [O\,{\sc iii}] and H$\alpha$+[N\,{\sc ii}] emission lines. 
These 3D views are computed from the VLT MUSE data cube using tailored {\sc python} routines such as {\sc Astropy}\footnote{{\sc Astropy} ia a Python package designed explicitly for Astronomy and maintained by the community \citep{Astropy2013}.} and the {\sc APLpy} tools, an open-source Python plotting package \citep{Aplpy2012}, and {\sc Mayavi} \citep{Mayavi2011}.
The first row from top to bottom of Fig.~\ref{fig:3D} shows a view of the these emission lines in the plane of the sky, where the depth corresponds to the velocity (or wavelength $\lambda$) dimension, whilst the other three rows show three rotated views. 
The first row thus shows direct images, yet the rainbow colour code provides additional information on the expansion velocity. 
The second and third row of Fig.~\ref{fig:3D} demonstrate that the H$\beta$ emission has indeed the simplest toroidal structure suggested by the integrated emission line map of Fig.~\ref{fig:muse_images}.
The toroidal structure is tilted with respect to the plane of the sky along a PA $\approx$5$^\circ$, with its eastern side approaching and the western side receding. 
Similarly, the central emission of the H$\alpha$ and the [N\,{\sc ii}] emission lines can be easily distinguished in the second row also as toroidal structures, in addition to a hint of extended (but blended) emission. 
Finally, the central emission of the [O\,{\sc iii}] line shows a toroidal structure, but its emission in the third row of Fig.~\ref{fig:3D} clearly reveals an open bipolar structure protruding ortogonally to the toroidal structure. 
This bipolar structure corresponds to the blue- and red-shifted clumps observed in the [O~{\sc iii}] channel maps both Eastwards and Westwards of the central star of T\,Pyx. 
The toroidal equatorial structure and the open bipolar lobes resemble a diabolo (this structure is also discussed later in this article and shown in Fig.\ref{fig:inc}).

\subsection{Multi-epoch images}
\label{subs.anal.exp}

\begin{figure*}
\begin{center} 
\includegraphics[angle=0,width=0.95\linewidth]{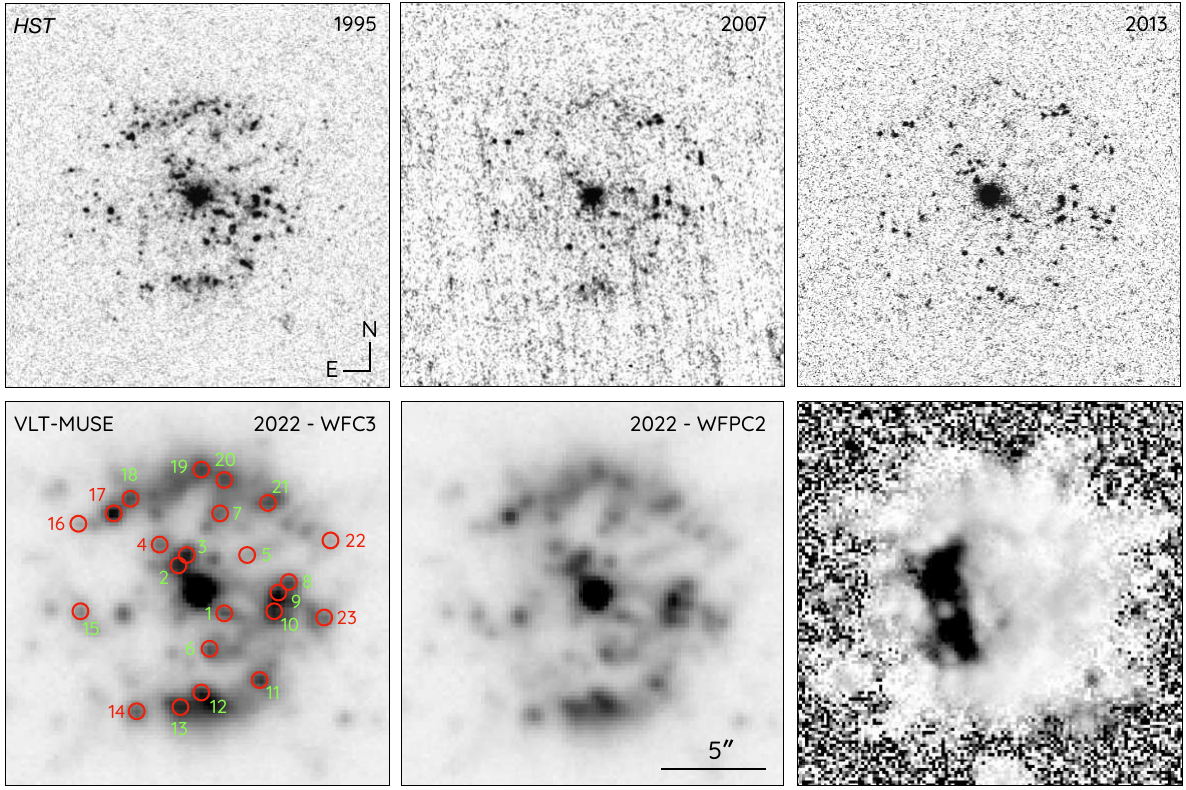}
\caption{
Nebular images of T~Pyx. 
Top panels: {\it HST} [N\,{\sc ii}] images obtained in 1995 and 2007 with the WFPC2 F658N filter and in 2013 with the WFC3/UVIS F658N filter. 
The bottom left and central panels show VLT MUSE images obtained by adopting the transmission curves of the WFPC2 and WFC3/UVIS F658N filters, respectively. 
The bottom right panel is a ratio of the two VLT MUSE images.
The field of view is the same in all panels. 
The bottom-left panel shows the knots identified in each multi-epoch image marked with red circles and red or green (for clarity) numbers. 
Their coordinates and radial distances are listed in Table~\ref{tab:coords}.
}
\label{fig:hstmuse}
\end{center} 
\end{figure*}

\begin{figure*}
\begin{center} 
\includegraphics[angle=0,width=0.95\linewidth]{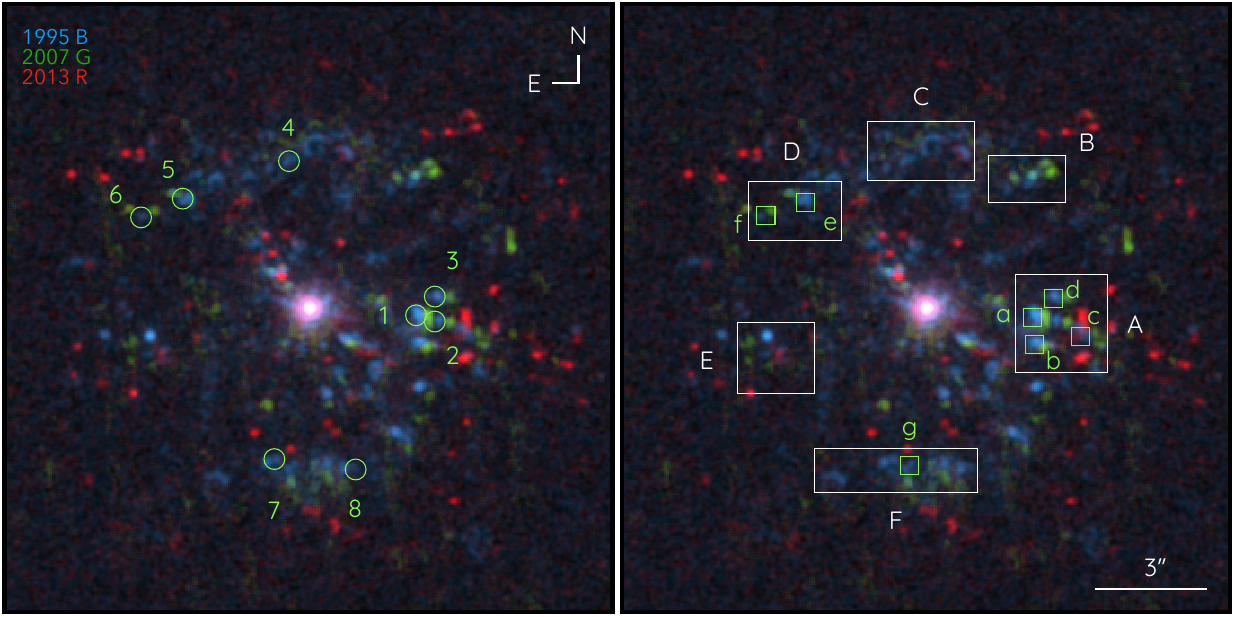}
\caption{
Multi-epoch {\it HST} [N\,{\sc ii}] images of T~Pyx. (top) RGB image composition from 1995, 2007 and 2013 data. The colour code as marked in the top-left corner of the left image is: Blue-1995, Green-2007 and Red-2013. The left-hand and right-hand panels shows the identification of individual features and regions used to estimate the expansion of the ejecta of T\,Pyx using the QM method. The squares and circles have an angular size of 0.42 and 0.41 (diameter) arcsec, respectively.}
\label{fig:HST}
\end{center} 
\end{figure*}

As noted in previous studies, the nova remnant of T\,Pyx presents a large number of knots mapped by [N~{\sc ii}] narrow-band images (see Section~\ref{sec:intro}). 
These are clearly seen in the {\it HST} WFPC2 F658N 1995.75, 2007.42 and WFC3/UVIS F658N 2013.75 images in the top panels of Fig.~\ref{fig:hstmuse}. 
An inspection of the colour-composite picture of the {\it HST} [N\,{\sc ii}] images of T\,Pyx obtained on 1995, 2007, and 2013 (Fig.~\ref{fig:HST}) reveals the evident angular expansion of this knotty structure. 
To estimate the present expansion of T\,Pyx and to investigate the dynamics of the ejected material as well as possible anisotropies, all \emph{HST} WFPC2 and WFC3/UVIS images and the VLT MUSE maps covering a time lapse from 1994.2 to 2022.0 (i.e., almost 28 years) will be used.
When comparing a pair of images of different epochs, the images are rebinned to the same pixel size, smoothed to have the same spatial resolution, and accurately aligned, which was performed using reference stars in the field of view (FoV) with standard {\sc iraf} \citep{1993ASPC...52..173T}\footnote{{\sc iraf} is distributed by the National Optical Astronomy Observatory, which is operated by the Association of Universities for Research in Astronomy (AURA) under a cooperative agreement with the National Science Foundation.} routines, and their surface brightness is scaled to have the same nebular peak value.

The WFPC2 F658N and WFC3/UVIS F658N filters transmit mainly the emission in the [N\,{\sc ii}] $\lambda$6584 line, but
the expansion velocity up to 1000~km~s$^{-1}$ of individual knots in T\,Pyx (see Fig.~\ref{fig:channel2} and \ref{fig:channel1}) anticipates a non-negligible H$\alpha$ contamination from the fastest receding material as well as a leakage of [N~{\sc ii}] $\lambda$6584 emission from the fastest approaching material. 
This is readily illustrated in Fig.~\ref{fig:trans} of Appendix~\ref{app:lines}.
Moreover the transmission curves of the WFPC2 F658N and WFC3/UVIS F658N filters are different, as noted in Sect.~\ref{sec:obs}.
To assess these effects, VLT MUSE emission maps have been obtained adopting the transmission profiles of the WFPC2 F658N and WFC3 F658N narrow-band filters, respectively. These images as well as a residual map of the VLT MUSE WFPC2-like and WFC-like F658N maps are presented in the bottom row of Fig.~$\ref{fig:hstmuse}$. The differences are negligible, but for an eastern region close to the central star, where H$\alpha$ and [N~{\sc ii}] emissions blend. This region will therefore be avoided in subsequent analyses. 

The quantified magnification (QM) method of minimization of residuals described by \cite{2019MNRAS.483.3773S} will be applied to single features (clumps) and global regions (SF and GF, respectively) on the pairs of \emph{HST} WFPC2 1994-2007 and 1995-2007 images to provide an accurate determination of their angular expansion rate. 
The QM method consists in the minimisation of residuals between images from different epochs. 
This procedure magnifies the earliest epoch image by a factor 1+$f$ before subtracting it from a more recent image. For example, for the 1994-1995 pair of images, the magnification factor is applied to the 1994 image before subtracting it from the 1995 image.
The optimal magnification factor is estimated by calculating the statistics in the difference image for each value of $f$ in regions around noticeable knots, minimising the dispersion, which is noted to decrease until a minimum value is reached to then start increasing again. 
The QM method can be applied to single features (SF-QM) and to global features or large regions (GF-QM) as performed in \citet{2019MNRAS.483.3773S} for the nova shell IPHASX\,J210204.7$+$471015.

After inspecting the expansion patterns exhibited by the clumps and filaments in the multi-epoch {\it HST} images of Fig.~\ref{fig:HST}, we defined several circular, square and rectangular regions encompassing different clumps and larger morphological features that were labelled with numbers and lower-case and capital letters (see Fig.~\ref{fig:HST}). 
The green circles in the left panel (labelled from {\it 1} to {\it 8}) were used to compute expansion patterns according to the SF-QM procedure on the 1994-1995 pair of images,  
the green squares in the right panel (labelled from {\it a} to {\it g}) were used in the SF-QM analysis of the 1994-2007 pair, 
the large rectangular regions (labelled from {\it A} to {\it F}) were used to study global expansion patterns for the 1994-2007 pair.

We note that the angular expansion of a series of knots of T\,Pyx was investigated by \citet{Schaefer2010} using the \emph{HST} WFPC2 1994.2, 1995.8 and 2007.4 images.  
We thus start here using these same \emph{HST} WFPC2 images, comparing the 1994-1995 and 1994-2007 pairs. 
Table~\ref{tab:obs} lists the magnification factors $f$ and angular expansion values and expansion rates for all selected regions illustrated in Fig.~\ref{fig:HST}. 
For the 1994-1995 image pair, the SF-QM procedure was used, whereas for the 1994-2007 image pair both SF-QM and GF-QM procedures were applied. 
A similar approach is questionable to the 1994-2013 and 1995-2013 pairs given the differences in the transmission curves of their filters illustrated in
Appendix~\ref{app:lines}. We note, none of our designations refer to same knots as used in \citet{Schaefer2010} and \citet{Shara+1997}.

\begin{table}
\caption{T\,Pyx expansion measurements. The labels corresponds to the regions marked on Fig.~\ref{fig:HST}.}
\label{slits}
\centering
\setlength{\tabcolsep}{0.1\tabcolsep}   
\label{tab:obs}
\begin{tabular}{ccccc} 
\hline
Feature &   Emission  &  Angular    & Expansion         & Magnification    \\
 label  &   line  &    Expansion    &  rate               &      $f$       \\
       &          &    (arcsec)     &  (arcsec yr$^{-1}$) &                \\
\hline
1994.16-1995.75 (SF-QM)\\
1 &  [N\,{\sc ii}] & 0.0421 & 0.0263 &  0.0134  \\
2 &  [N\,{\sc ii}] & 0.0440 & 0.0275 &  0.0120  \\
3 &  [N\,{\sc ii}] & 0.0398 & 0.0248 &  0.0109  \\
4 &  [N\,{\sc ii}] & 0.0519 & 0.0324 &  0.0126  \\
5 &  [N\,{\sc ii}] & 0.0560 & 0.0350 &  0.0123  \\
6 &  [N\,{\sc ii}] & 0.0671 & 0.0419 &  0.0130  \\
7 &  [N\,{\sc ii}] & 0.0782 & 0.0488 &  0.0184  \\
8 &  [N\,{\sc ii}] & 0.0773 & 0.0483 &  0.0172  \\
\hline
1994.16-2007.42 (SF-QM)\\
a &  [N\,{\sc ii}] & 0.2890 & 0.0217 &  0.0918  \\
b &  [N\,{\sc ii}] & 0.3096 & 0.0232 &  0.0939  \\
c &  [N\,{\sc ii}] & 0.4184 & 0.0314 &  0.0924  \\
d &  [N\,{\sc ii}] & 0.3718 & 0.0279 &  0.0959  \\
e &  [N\,{\sc ii}] & 0.3795 & 0.0285 &  0.0849  \\
f &  [N\,{\sc ii}] & 0.4647 & 0.0349 &  0.0905  \\
g &  [N\,{\sc ii}] & 0.4618 & 0.0347 &  0.1048  \\
\hline
1994.16-2007.42 (GF-QM)\\
A &  [N\,{\sc ii}] & 0.3801 & 0.0285 &  0.0977  \\
B &  [N\,{\sc ii}] & 0.4872 & 0.0366 &  0.1020  \\
C &  [N\,{\sc ii}] & 0.4856 & 0.0365 &  0.1000  \\
D &  [N\,{\sc ii}] & 0.4269 & 0.0321 &  0.0901  \\
E &  [N\,{\sc ii}] & 0.4841 & 0.0369 &  0.1026  \\
F &  [N\,{\sc ii}] & 0.5696 & 0.0428 &  0.1207  \\
\hline
\end{tabular}
\end{table}

\begin{figure}
\includegraphics[width=1.0\linewidth]{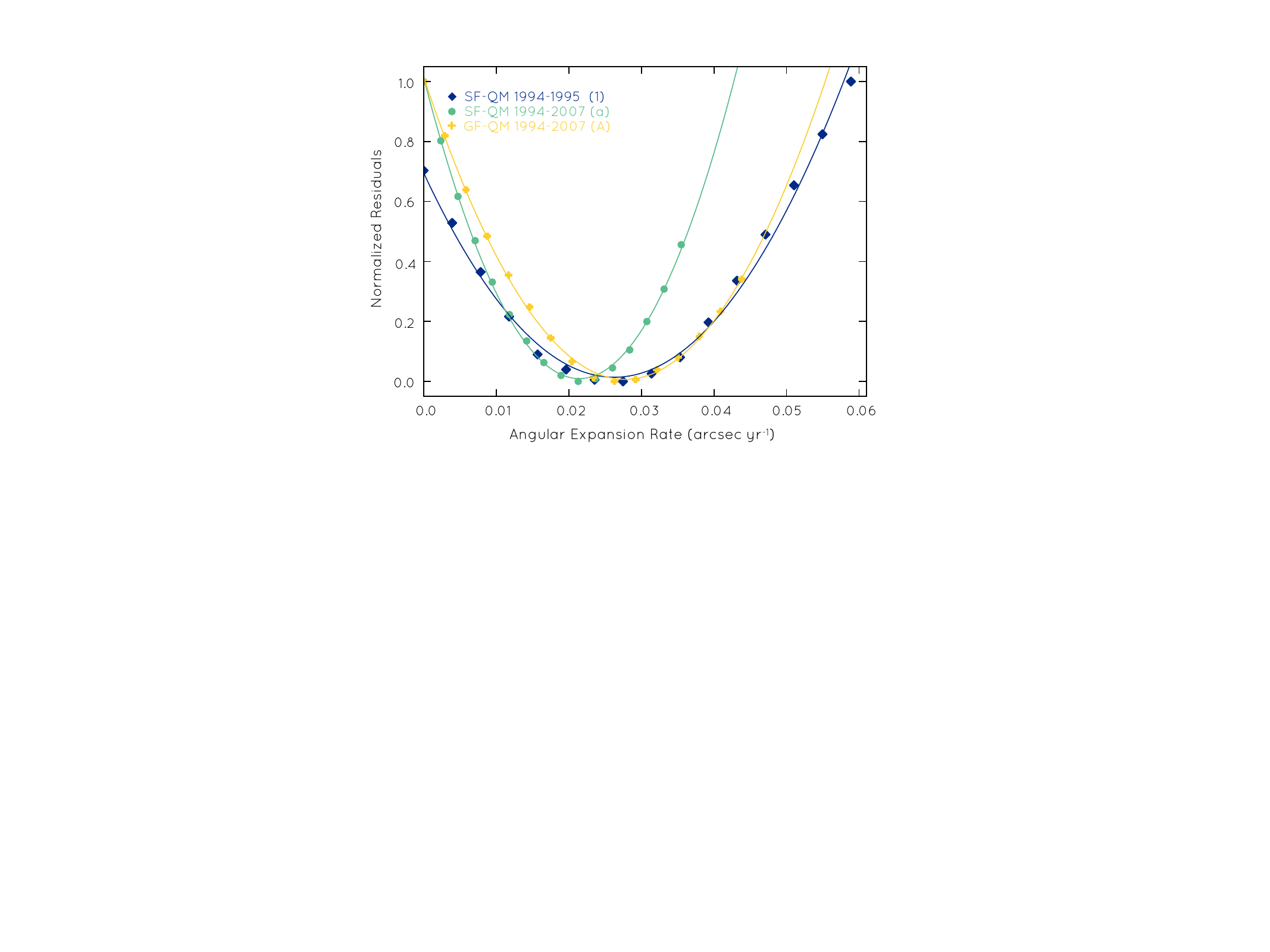}
\caption{
Normalised residuals of apertures \emph{1}, \emph{a} and \emph{A} for the pairs of image epochs labelled and best least-squares fits using the QM methods described in the text. The expansion rate is determined from the minima of the least-squares fits.
}
\label{fig:fits}
\end{figure}

\begin{figure}
\includegraphics[width=1.0\linewidth]{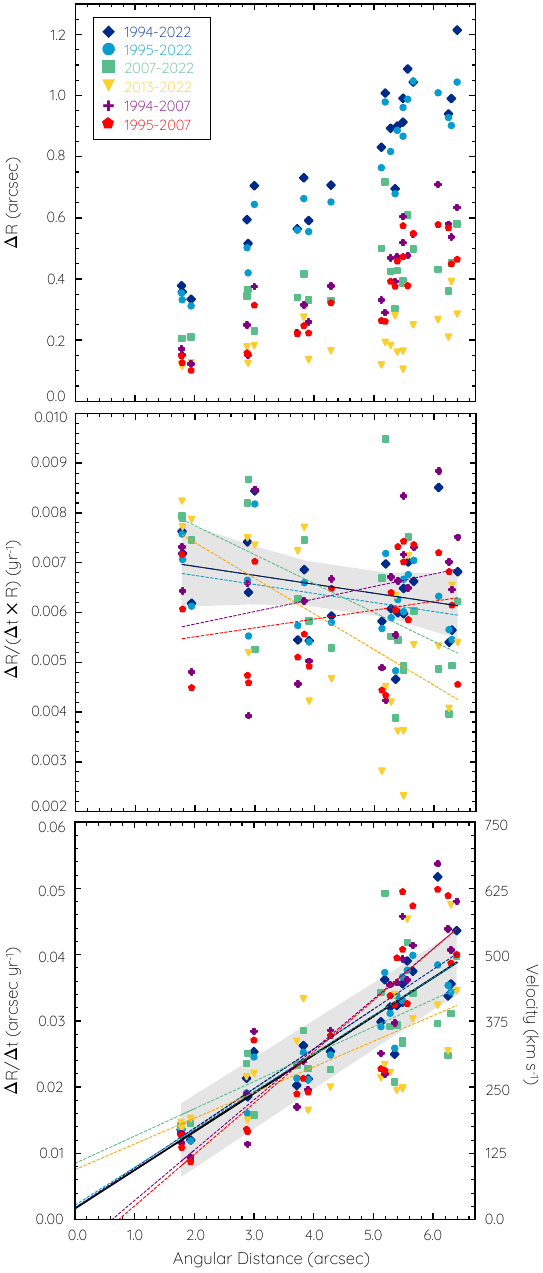}
\vspace*{-0.5cm}
\caption{
Radial variation of the angular expansion of T\,Pyx. 
(top) Radial variation of the angular expansion $\Delta R$ of features marked in Fig.~\ref{fig:HST} for pairs of images obtained at different epochs as labeled.  
(middle) Same as top panel for $\Delta R/ (\Delta t \times R)$ with linear fits to each pair of images. 
A constant value would imply a homologous expansion.  
(bottom) Same as the panel above for the expansion rate $\Delta R/\Delta t$. 
The solid black line is a joint fit to all data points and the grey-shaded areas represent the 95 per cent confidence band.} 
\label{fig:radial}
\end{figure}

A direct comparison of the expansion rates derived by the SF-QM and GF-QM procedures is presented in Fig.~\ref{fig:fits} using clumps and filaments of T\,Pyx in the same vicinity, particularly the features \emph{1}, \emph{a}, and \emph{A} labelled in Fig.~\ref{fig:HST}.  
The best-fit minimisation curve derived from feature \emph{1} (blue line), using the image pair with the shortest 1994.2-1995.8 time-lapse, is notably broader than those obtained from features \emph{a} (green line) and \emph{A} (orange line) obtained from the image pair 1994.2-2007.4 with a longer time-lapse. 
This reflects the larger uncertainty of the fits to the 1994.2-1995.8 image pairs with the shortest time-lapse, which is in line with the larger dispersion of the values associated with these fits in Table~\ref{slits}. 
On the other hand, the expansion rates derived using the SF-QM procedure for feature \emph{a} (0\farcs0217~yr$^{-1}$, green line) and the GF-QM procedure for feature \emph{A} (0\farcs0285~yr$^{-1}$, orange line) differ as the aperture of feature \emph{A} is much larger than that of feature \emph{a}.  
Indeed the averaged SF-QM angular expansion rates of features \emph{a}, \emph{b}, \emph{c}, and \emph{d} included in feature \emph{A} ($\approx$0\farcs026~yr$^{-1}$) is very similar to that of feature \emph{A} (0\farcs0285~yr$^{-1}$).

To quantify the expansion of the knots with respect to the VLT MUSE maps, the different multi-epoch images were compared using the task {\sc kshell}\footnote{This tool allows exploring astronomical data cubes with expansive shells, estimating their radius directly and generating line profiles in the direction to be analyzed, as well as other features.} from the KARMA\footnote{\url{https://www.atnf.csiro.au/computing/software/karma}} package \citep{Gooch1996}.  
This task estimates the radial distance of the knots to the central star on radial profiles extracted along chosen directions to include notable knots at all epochs.
The radial profiles extracted from the corresponding VLT MUSE WFPC2-like and WFC3-like [N~{\sc ii}] maps were then shifted, and their difference with respect to the profiles from the respective \emph{HST} WFPC2 and WFC3/UVIS images was calculated.  
The best shift (and thus the angular expansion $\Delta R$) is computed as the one that minimizes the difference between these two spatial profiles at the location of the individual knots and features marked in Fig.~\ref{fig:hstmuse}.
We remark that the VLT MUSE maps have been created from the MUSE datacube taking into account the transmission curves of the WFPC2 and WFC3/UVIS F658N filters, which are illustrated in Fig. \ref{fig:trans}. 
The VLT MUSE WFPC2-like F658N map is then compared to the \emph{HST} WFPC2 F658N 1994.2, 1995.8 and 2007.4 images, whereas the VLT MUSE WFC3-like F658N map is accordingly compared to the \emph{HST} WFC3/UVIS F658N image. 
It is convenient to remark as well that all \emph{HST} images were scaled to the pixel size of the VLT MUSE data (i.e., 0.2$\times$0.2~arcsec$^{2}$) and smoothed with a Gaussian 2D kernel to match its spatial resolution of 1.0~arcsec.

The dependence of the angular expansion $\Delta R$ and expansion rate $\Delta R/\Delta t$ with angular distance $R$ is illustrated in Fig.~\ref{fig:radial}.  
The angular expansion $\Delta R$ of each feature (see left panels of Fig.~\ref{fig:hstmuse} and \ref{fig:HST}), presented in the top panel of Fig.~\ref{fig:radial}, exhibits a clear increase with radial distance for all pairs of images.  
This is consistent with the homologous expansion of the ejecta of T\,Pyx described by \citet{Schaefer2010}, which specifically implies a 
linear dependence of the angular expansion (or the angular expansion rate $\dot R$) with the radial distance 
\begin{equation}
    \Delta R \propto R \times \Delta t, \;\;\; {\rm or} \;\;\; \frac{\Delta R}{\Delta t} = \dot R \propto R.
\end{equation}

This relationship is checked in the middle panel of Fig.~\ref{fig:radial}, where $\Delta R/\Delta t / R$ (or $\dot R/R$) is plotted against $R$, together with linear fits to the data sets of each pair of images.  
Individual fits and the parameters of the fits are given in Fig.~\ref{fig:acel} and Tab.~C1 of Appendix~\ref{app:indfits}.  
The averaged value $\Delta R/\Delta t / R$ of all data points in these plots is 0.0062$\pm$0.0003 yr$^{-1}$. A homologous expansion would be revealed by a constant value of $\Delta R/\Delta t / R$, which is the case for the 1994-2007, 1994-2022, 1995-2007, and 1995-2022 pairs of images. Otherwise the linear fits to the 2007-2022 and 2013-2022 pairs of images have negative slopes at 3.2 and 3.6 $\sigma$ confidence levels, respectively. This would imply that the expansion in the period from 2007 to 2022 is not homologous, but it is proportionally larger for knots closer to the central star.

These trends are illustrated as well in the bottom panel of Fig.~\ref{fig:radial}, which plots $\Delta R/\Delta t$ vs.\ $R$, together with linear fits to the data sets of each pair of images, and a joint linear fit to all data points (black line). 
Individual fits and the parameters of the fits are given in Fig.~\ref{fig:time} and Tab.~C1 of Appendix~\ref{app:indfits}.  
The linear fits show a consistent decrease of the slope from the oldest 1994-2007 and 1995-2007 to the newest 2007-2022 and 2013-2022 pairs of images.  
It should also be noted that all linear fits are consistent to go across the origin of this diagram within 1-$\sigma$ uncertainty of the fit, but those of the latest 2007-2022 and 2013-2022 image pairs.

\section{Discussion}
\label{sec:discussion}

\subsection{The 3D structure of T Pyx}
\label{sec:3Dstructure}

The VLT MUSE line-integrated maps of T\,Pyx unveil diffuse, more extended nebular emission than the clumpy nova shell disclosed by \emph{HST} images.  
The deep VLT MUSE maps and sharp \emph{HST} images therefore provide a complementary view of this nova remnant.  
Moreover the tomographic capabilities of the VLT MUSE observations uniquely disclose its true 3D spatial structure through its morpho-kinematic signature.

The brightest component of the nova remnant consists of a tilted clumpy toroidal structure, best seen in the H$\beta$ PPV diagrams in Fig.~\ref{fig:3D}. 
The aspect ratio of the projected image of this toroidal structure and its radial expansion velocity along the minor axis imply an inclination angle of its symmetry axis with the line of sight $\approx$51$^\circ$ and an expansion velocity $\approx$470 km~s$^{-1}$.  
The inclination of this toroidal structure can also be derived comparing the observed radial velocity $\approx$370 km~s$^{-1}$ along its minor axis with the tangential velocity of features \emph{a} and \emph{b} at this location (see right panel of Fig.~\ref{fig:HST}) under the only assumption of radial expansion. 
These have angular expansion rates of 0.0217 and 0.0232 arcsec~yr$^{-1}$, respectively (Table~\ref{tab:obs}), that correspond to an averaged tangential velocity $\approx280$ km~s$^{-1}$ at the distance of 2.6 kpc.  
The comparison of radial and tangential velocities results in an inclination angle of the symmetry axis of the toroid with the line of sight of 53$^\circ$, and an expansion velocity $\approx 460$ km~s$^{-1}$. 
Both the interpretation of this structure as an expanding circular toroid and its inclination angle and expansion velocity are similar to those ($i = 64^\circ$$^{+21}_{-14}$, $V_{\rm exp} = 470^{+80}_{-70}$ km~s$^{-1}$) recently reported by \citet{Izzo2023} using the same VLT MUSE dataset. 
At the \emph{Gaia} distance of $2.6^{+0.4}_{-0.2}$ kpc and under the assumption of free expansion, the angular distance and expansion velocity of the toroidal structure imply an age of $150\pm36$ yr and an ejection time circa $1873\pm36$. 
These are quite similar to the results proposed by \citet{Schaefer2010}, for a time of the event circa the year of 1866.

The H$\beta$ toroidal structure seems to be the waist where a pair of bowl-shaped open structures join, best seen in the [O~{\sc iii}] PPV diagrams in Fig.~\ref{fig:3D}.
To further endorse this statement, we show in Fig.~\ref{fig:inc} 3D visualisations from different lines of sight of the H$\beta$ and [O\,{\sc iii}] emission lines (i.e., those that are isolated) extracted directly from the MUSE data cubes. 
The 3D visualisations in Fig.~\ref{fig:inc} clearly reveals the advantage of the analysis of spectral cubes, which provides the opportunity to reveal the morphology without geometrical assumptions, deproject the structure, and observe it along any line of sight. 
The bipolar morphology of the nova ejecta of T\,Pyx is obvious in the 3D visualisation of the [O\,{\sc iii}] emission line, more clearly highlighted by the mesh pattern built with a bipolar {\sc shape} \citep{Steffen2011} model.  
The inclination angle of its symmetry axis with the line of sight is $\approx56^\circ$, i.e., it is orthogonal to the H$\beta$ toroidal structure.

It is noted the presence of HVCs in the 3D visualisation of the [O\,{\sc iii}] emission line.  
These were noted as well in the He~{\sc ii} and [O\,{\sc iii}] emission line profiles (Fig.~\ref{fig:muse_lines}-top) and in the [O~{\sc iii}] channel maps (Fig.~\ref{fig:channel2}). 
Unlike the fast knots in the close vicinity of the central star of T\,Pyx reported by \citet{Izzo2023}, which are associated to the most recent 2011 outburst of T\,Pyx, the HVCs noted here can be ascribed to the bipolar structure delineated by the mesh pattern in the bottom panels of Fig.~\ref{fig:inc}.  
Thus this HVCs can be associated to the same ejection that produced the large-scale knotty structure of the T\,Pyx remnant in the ninetieth century.  

\subsection{The angular expansion of T~Pyx}

The angular expansion of the remnant associated with the recurrent nova T\,Pyx is revealed by comparing multi-epoch \emph{HST} WFPC2 and WFC3 narrow-band images (e.g., the \emph{HST} picture presented in Fig~\ref{fig:HST}) and VLT MUSE emission line maps spanning from 1994.2 up to 2022.0 presented in Section~\ref{subs.anal.exp}.  
The dependence of the angular expansion and expansion rate with the radial distance of individual knots and nebular features is shown in Fig.~\ref{fig:radial} jointly for all image pairs and individually in Figs.~\ref{fig:acel} and \ref{fig:time} of Appendix~\ref{app:indfits}.

The angular expansion of the remnant of T\,Pyx was already proven by \citet{Schaefer2010} using multi-epoch \emph{HST} WFPC2 images obtained on 1994.2, 1995.8, and 2007.4. 
They measured expansion velocities ranging from 500 to 700 km~s$^{-1}$ at their assumed distance of 3.5 kpc and concluded that the observed fractional expansion of the knots remains unchanged, indicating minimal deceleration caused by the interstellar or circumstellar media.
These results are recovered by the analysis of these pairs of \emph{HST} WFPC2 images in Section~\ref{subs.anal.exp}, with tangential expansion velocities $\approx$550 km~s$^{-1}$ (see the panels corresponding to the 1994-2007 and 1995-2007 pairs of images of Fig.~\ref{fig:time}), in agreement with \citet{Schaefer2010}'s results given the smaller 2.6 kpc distance adopted here. 
The averaged expansion of 0.0062$\pm$0.0003 yr$^{-1}$ derived in Section~\ref{subs.anal.exp} implies an age of 161$\pm$8 yr and thus an ejection time circa $1861\pm8$ AD. 
These are consistent with those derived from the spatio-kinematics of the clumpy toroidal structure derived in Section~\ref{sec:3Dstructure} and those presented by \citet{Schaefer2010}.

The comparison of later pairs of images, particularly those of 2007-2022 and 2013-2022, suggests that the slope of the best linear fits of the expansion of these pairs of images is negative.  
This would imply that the expansion pattern is not homologous, but the fractional expansion is larger for knots closer to the central star. 
The outermost knots are either slowing down, which was refuted by \citet{Schaefer2010}, or the innermost knots are being sped up.

The implications for the effects of the recurrent outbursts of T\,Pyx on the expansion of its nebular remnant would be important, and deserve the acquisition of present day \emph{HST} WFC3/UVIS images to be compared with the available 2013 image.  
The homologous expansion during the 1994 to 2007 period as compared to that during the 2007 to 2022 period requires  dynamical effects starting at some moment during 2007 and 2013 that sped up the innermost knots. 
This can be attributed to the outburst experienced by T\,Pyx on 2011 April 14.  
Let's assume that the angular expansion $\Delta R$ of a knot at distance $R$ changed then from homologous
\begin{equation}
    \Delta R_{\rm hom} = \dot R_{\rm hom} \times R \times \Delta t, 
\end{equation}
to include an additional term $\delta R$ 
\begin{equation}
    \Delta R_{\rm obs} = \Delta R_{\rm hom} + \delta R.
\end{equation}
If this additional expansion term $\delta R$ were to be associated with the 2011 nova outburst, we can expect it to decrease with radial distance, as the effect of the outburst needs to propagate through the nebular remnant.  
Even if it is assumed to be a constant, then the observed expansion rate 
\begin{equation}
    \dot R_{\rm obs} = \frac{\Delta R_{\rm obs}}{\Delta t \times R} = \dot R_{\rm hom} + \frac{\delta R}{\Delta t \times R} = \dot R_{\rm hom} + \delta \dot R
\end{equation}
also includes an additional term $\delta \dot R$ that is proportional to $R^{-1}$, thus increasing at low radial distances and producing a negative slope. 
This relationship also reveals that $\delta \dot R$ is proportional to $\Delta t^{-1}$.  
The time-lapse between pairs of images thus dilutes the effects of the additional non-homologous expansion, making it smaller (or even unnoticed) for the 1994-2022 and 1995-2022 pairs of images, and increasing gradually for the 2007-2022 and most notably for the 2013-2022 pairs of images.  

The 3D physical structure of T\,Pyx was described in Section~\ref{sec:3Dstructure} to be similar in shape to a tilted diabolo.  
Knots apparently closer to the central star may actually be further away than knots projected far. 
If the additional $\delta R$ expansion term were to depend on the linear distance to the central star, this relation does not propagate to the projected radial distances considered in the equations above, introducing a dispersion on the correlation between radial distance and expansion.  
Indeed the correlation coefficients of the best linear fits to the 2007-2022 and 2013-2022 pairs of images in Table \ref{fig:time} are actually smaller, $\approx$0.65, than the same coefficients for the other pairs of images, typically $\approx$0.90.

The closest nebular structure  to the central star of T\,Pyx here analyzed is the toroidal waist of the diabolo structure.  
Its radius of 5.6~arcsec correspond to a physical distance of 0.07 pc at a distance of 2.6 kpc.  
The material ejected during the April 2011 outburst has been found to have an expansion velocity 2200-2400 km~s$^{-1}$ \citep{2011ATel.3289....1N,2012PASJ...64L...9I}.
This material has thus travelled $\simeq$0.025 pc in the 10.7 yr period from the outburst to January 2022 when the VLT MUSE data were acquired.  
Therefore it has had no time to reach the toroidal waist of the diabolo structure, but its expected angular size $\simeq$2 arcsec is revealed in high-resolution VLT MUSE observations using the CLT AO module and MUSE NFM \citep{Izzo2023}.

Therefore the acceleration of the inner knots of T\,Pyx cannot be attributed to their interaction with the most recent ejecta.  
\citet{Shara+1997} and \citet{Schaefer2010} noted that the knots of T\,Pyx brighten and fade with time.  
Among other possible explanations, the flash ionization of the knots following the outburst, moving the ionization front inwards them and thus further away from the central ionizing source, can certainly mimic an increased expansion rate.  
Alternatively the present knots in the nebular remnant of T\,Pyx can interact collisionally with material from previous outbursts, not specifically the last one.

To conclude, we remark that the apparent acceleration of the knots after the 2011 outburst of T\,Pyx and the notorious recurrence of this nova might have produced similar effects before.  
This casts doubts onto the estimate of expansion age so far derived, implying that it can be shorter than thought (and its outburst time closer in time). 
The detailed comparison of \emph{HST} WFPC2 images previous to the 2011 outburst needs to be revised with post-outburst images, comparing the 2013 \emph{HST} WFC3/UVIS image with a present day image.

\subsection{Ionised mass and kinematics}

The H$\alpha$ line cannot be used to estimate the ionised mass of the nebular remnant of T\,Pyx as typically done for other nebular objects \citep[see, e.g.,][]{Santamaria2022,Rechy2022} given that it is blended with the [N\,{\sc ii}] $\lambda\lambda$6548,6584 emission lines. 
Instead, the H$\beta$ line will be used for this purpose.

The total observed H$\beta$ flux derived from MUSE observations is $3.8\times10^{-14}$ erg cm$^{-2}$ s$^{-1}$, which after correcting for an extinction of 
$c$(H$\beta$)=0.7 derived from the VLT MUSE spectrum of a number of clumps free from the [N\,{\sc ii}] emission lines, results in an estimated intrinsic H$\beta$ flux of $F_\mathrm{H\beta}$=6.4$\times10^{-14}$ erg cm$^{-2}$ s$^{-1}$, which corresponds to an H$\alpha$ flux of $F_{\rm H\alpha}=1.83\times10^{-13}$~erg cm$^{-2}$ s$^{-1}$ adopting a theoretical ratio of H$\alpha$/H$\beta$=2.85 for an ionised ($T=10^{4}$ K) H-rich gas \citep{2006agna.book.....O}.

The electron density ($n_\mathrm{e}$) can be calculated from $F_{\rm H\alpha}$ using the expression \citep[see][]{1968IAUS...34..162B}
\begin{equation}
    n_\mathrm{e} = 1.2 \sqrt{\frac{4 \pi d^2 F_\mathrm{H\alpha}}{\varepsilon V C_\mathrm{H\alpha}}},
\end{equation}
\noindent where 
$C_\mathrm{H\alpha}$ is the emission coefficient of the H$\alpha$ line with a value of $4\times10^{-25}$ erg~cm$^{-3}$~s$^{-1}$, $V$ is the emitting volume and $\varepsilon$ is the filling factor which can be described as $\varepsilon=a\times b$, where $a$ and $b$ are the macroscopic and microscopic components, respectively \citep[see details in][]{Santamaria2022}. This equation assumes that the ratio of the electron density over the proton density ($n_\mathrm{e}$/$n_\mathrm{p}$) is equal to 1.2. 
The volume of the emitting region can be calculated adopting a toroidal structure with total radius of $R$=5.6~arcsec (=0.071 pc) and an angular radial width of $r$=0.5~arcsec (=$6.3\times10^{-3}$ pc), i.e., $V=2\pi^{2} R r^{2}  \approx 1.6\times10^{51}$ cm$^{-3}$. 
Therefore the average electron density is estimated to be $n_\mathrm{e}\approx570 \times \varepsilon^{-1/2}$ cm$^{-3}$.

The total ionised mass of the remnant can then be calculated as:
\begin{equation}
    M_\mathrm{shell} = \mu m_{\rm p} \sqrt{\frac{4 \pi d^2 \varepsilon V F_\mathrm{H\alpha}}{C_\mathrm{H\alpha}}},
\end{equation}
\noindent where $\mu$ is the mean molecular weight, which can be assumed to be 1.44 for a He/H solar ratio. This translates into a total ionised mass of $M_\mathrm{ion}=9.3\times10^{-4} \times \varepsilon^{1/2}$ M$_\odot$.

Alternatively the total mass of the nebula can be estimated following this same procedure but for each spaxel of the MUSE data. 
Thus for each volume element the H$\beta$ flux is measured and its corresponding $n_\mathrm{e}$ and mass computed. 
This way a total mass of $7.5\times10^{-4} \times b^{1/2}$~M$_\odot$ is estimated. 
A comparison of this mass and the average mass computed above implies a value for the filling factor component $a$ of 0.8.

\cite{2008Selvelli} estimated an ejected mass of 10$^{-4}$ to 10$^{-5}$ M$_\odot$ during the 1967 eruption based on the long duration of the optically thick phase in the optical light curve, but with significant uncertainties. \cite{Schaefer2010} found that some knots have turned on in the 2007 image, while others faded 1.5 mag since the 1995 image. 
This late turn-on of many knots shows that collisional shocks must power them. With corrections for ignition and knot fading, they estimate a total ejected mass of $3\times10^{-5}$ M$_\odot$.  

It is important to bear in mind that using H$\beta$ imposes a lower limit in the total ionised mass of the nova remnant, as the H$\alpha$ emission, being brighter, detects further extended emission. 
By comparing the H$\beta$ channel maps with other emission lines, for example that of the [O\,{\sc iii}] 5007~\AA\, emission, it is obvious that some extended emission is not detected by the former. However, we note that most of the mass should be contained within the central toroidal structure traced by the H$\beta$ line \cite[see the case of RR Pic;][]{2023Celedon}.  
Finally, we note that \citet{Izzo2023} estimated a mass for the delayed 2011 ejection of $<(3\pm1)\times10^{-6}$~M$_\odot$. For comparison, we can estimate an averaged ejected mass of 1.25$\times10^{-4}$~M$_\odot$ for the six nova events experienced by T Pyx. If the averaged value is indeed larger than that estimated only for the 2011 ejection by \citet{Izzo2023}, this might suggest that the ejected mass per nova event might not be constant, and neither  the time-lapse between outbursts.

The total kinetic energy of the remnant, which can be defined as $E_\mathrm{k} = 1/2 ~M_\mathrm{ion}~v_\mathrm{exp}^{2}$, is then estimated to be about $1.6\times10^{45}$~erg for an expansion velocity of $v_\mathrm{exp}=$460~km~s$^{-1}$. This value is about an order of magnitude above the kinematic energies reported for other nova remnants \citep[e.g.,][]{Santamaria2020,Santamaria2022}. However, the total kinetic energy of T Pyx should include the energy deposited during the six reported ejections.

Adopting a density for the interstellar medium around T\,Pyx of 1~cm$^{-3}$, the displaced mass  
can be estimated to be [2--3]$\times10^{-5}$~M$_\odot$ for an hourglass or a two-ellipsoidal morphology. 
This can be reduced to less than 0.5$\times10^{-5}$~M$_\odot$ if the high height of T\,Pyx above the Galactic Plane is considered. 
The latter is consistent with findings that the nova remnant is still in free expansion phase, which is also found for other nova remnants \citep[see][]{Santamaria2022}.

\section{Summary}
\label{sec:summary}

This paper presents an analysis of archival VLT MUSE IFS observations and multi-epoch \emph{HST} WFPC2 and WFC3/UVIS narrow-band images of the remnant associated with the recurrent nova T\,Pyx.  
The true 3D structure of T\,Pyx revealed by the VLT MUSE tomographic view is found to be bipolar, with an open diabolo-shaped structure best seen in [O\,{\sc iii}] emission lines and a knotty waist best revealed in the H$\beta$ emission line.  
The toroidal structure expands at $\approx$460 km~s$^{-1}$ and its symmetry axis is tilted by $\approx52^\circ$ with the line of sight.

The comparison of multi-epoch \emph{HST} WFPC2 and WFC3/UVIS narrow-band images and VLT MUSE emission line maps of T\,Pyx discloses the proper motion on the plane of the sky of individual knots and nebular features. 
The nebular remnant of T\,Pyx presents a homologous expansion pattern in the time period from 1994.2 to 2007.4, but the expansion rate of the inner knots seems to have increased at least since 2013.8, which can be most likely associated with its April 2011 outburst. 
The increased expansion rate is unlikely to result from the dynamical interaction of the recent 2011 outburst with the previous ejecta, but it can be associated with enhanced ionization or collisional interactions with previous outbursts different to the last one.

At the \emph{Gaia} distance of 2.6 kpc, the expanding toroidal structure has a kinematical age of $150\pm36$ yr.  
The average slope of the best linear fits to the angular expansion between different epochs is $0.0062\pm0.0003$ yr$^{-1}$, then resulting in an expansion age of $161\pm8$ yr.  
The corresponding times of the outburst according to these two estimates are $1873\pm36$ and $1861\pm8$, respectively.  
These agree among them as well with the 1866 event proposed by \citet{Schaefer2010}, although it is noted that the hypothesis of homologous ballistic expansion is questioned as the recurrent outbursts of T\,Pyx might have had non negligible effects on the dynamics of the mid-ninetieth century ejecta.

\section*{Acknowledgements}

The authors thank the anonymous referee for comments and suggestions that improved the presentation of the present work. E.S.\ thanks UNAM DGAPA for a postdoctoral fellowship. 
J.A.T.\ thanks support from the UNAM DGAPA PAPIIT project IN102324. 
M.A.G.\ acknowledges financial support from grants CEX2021-001131-S funded by MCIN/AEI/10.13039/501100011033 and PID2022-142925NB-I00  from the Spanish Ministerio de Ciencia, Innovaci\'on y Universidades (MCIU) cofunded with FEDER funds.
G.R-L acknowledges support from CONACyT grant 263373. L.S.\ thanks support from the UNAM DGAPA PAPIIT project IN110122.
This work has made an extensive use of NASA’s Astrophysics Data System (ADS).

\section*{DATA AVAILABILITY}

The data used in this article can be found in public archives. 
The processed VLT MUSE data are available at the ESO Archive Science Portal (\url{https://archive.eso.org/scienceportal/home}) and the {\it HST} data can be found in the Hubble Legacy Archive (\url{https://hla.stsci.edu/}).









\appendix

\section{H$\alpha$, H$\beta$ and [N\,{\sc ii}] emission line profiles}
\label{app:over}

\begin{figure}
\includegraphics[width=0.9\linewidth]{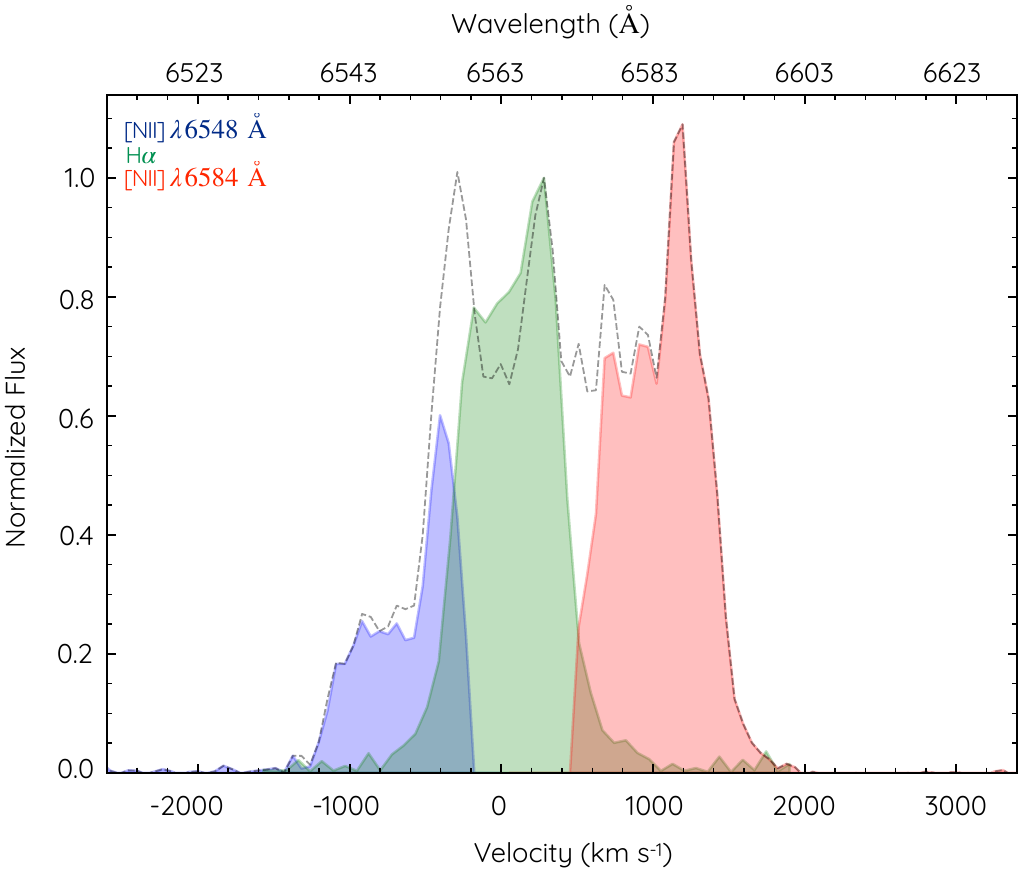}
\caption{
Difference between H$\alpha$+[N\,{\sc ii}] and H$\beta$ normalized line profiles to show possible isolated [N\,{\sc ii}] $\lambda$6548 (blue), H$\alpha$ (green) and [N\,{\sc ii}] $\lambda$6584 (red) line profiles. The velocity axis is referred to the H$\alpha$ line.}

\label{fig:subs}
\end{figure}

The remarkable blending between the H$\alpha$ and [N~{\sc ii}] nebular emission line profiles is highlighted by the normalized H$\beta$ emission line profile overplotted in Fig.~\ref{fig:muse_lines}, as the H$\alpha$ and H$\beta$ nebular line profiles are consistent with each other. 
The normalized H$\beta$ profile subtracted from the nebular line profile reveals the possible clean profiles of the [N~{\sc ii}] emission lines as shown in Fig.~\ref{fig:subs}. 
Apparent differences between the [N~{\sc ii}] $\lambda$6548 and [N~{\sc ii}] $\lambda$6584 emission profiles, such as the observed one-to-two ratio between their peak profiles, instead of the theoretical one-to-three ratio, testify the imperfect recovery of their line profiles. 

\section{Transmission curves of the {\it HST} F658N filter}
\label{app:lines}

\begin{figure}
\begin{center}
\includegraphics[width=1.0\linewidth]{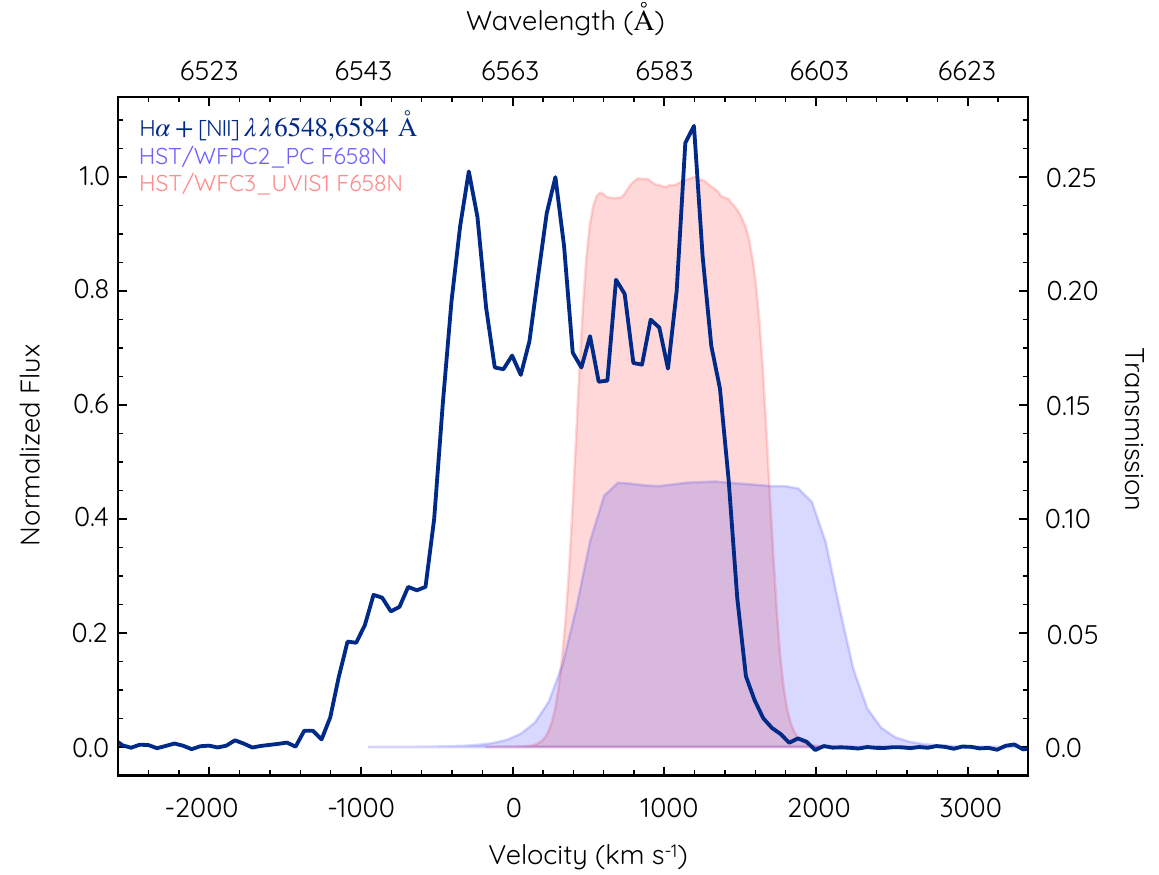}
\caption{
Transmission curves of the {\it HST} WFPC2 F658N (blue) and {\it HST} WFC3/UVIS F658N (red) filters. 
The blended H$\alpha$+[N\,{\sc ii}] profile is also shown. The velocity axis is referred to the H$\alpha$ line.
}
\label{fig:trans}
\end{center}
\end{figure}

In order to demonstrate the effects of the F658N filter of the {\it HST} WFPC2 and WFC3/UVIS instruments we show in Fig.~\ref{fig:trans} their transmission curves in comparison with the blended H$\alpha$+[N\,{\sc ii}] line profile. 
The figure shows that the transmission of the WFPC2 F658N filter includes emission from both the H$\alpha$ and [N\,{\sc ii}] $\lambda$6584 emission lines, as does the WFC3/UVIS F658N filter to a lesser extent. 

\section{Individual fits}
\label{app:indfits}

In this appendix we present the details to the expansion patterns obtained by comparing different pairs of images of T Pyx. The characteristics of the 23 selected clumps for this analysis are presented in Table~\ref{tab:coords1}.

\begin{figure*}
\begin{center}
\includegraphics[width=0.8\linewidth]{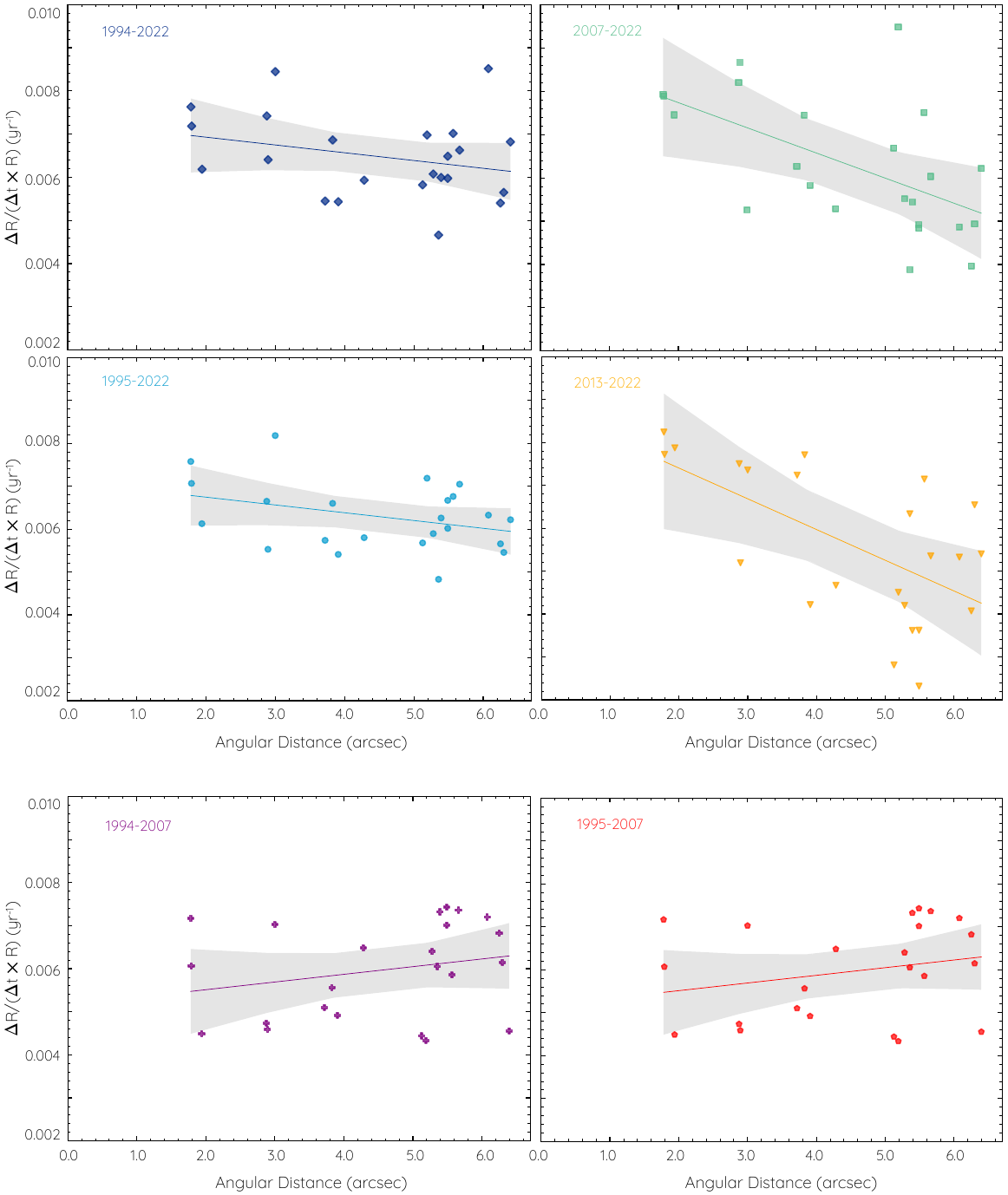}
\caption{Individual fits from middle panel Fig.~\ref{fig:radial}. 
The span of both the X and Y axes are respectively identical in all panels to allow a fair comparison of the different fits.}
\label{fig:acel}
\end{center}
\end{figure*}

\begin{figure*}
\begin{center}
\includegraphics[width=0.85\linewidth]{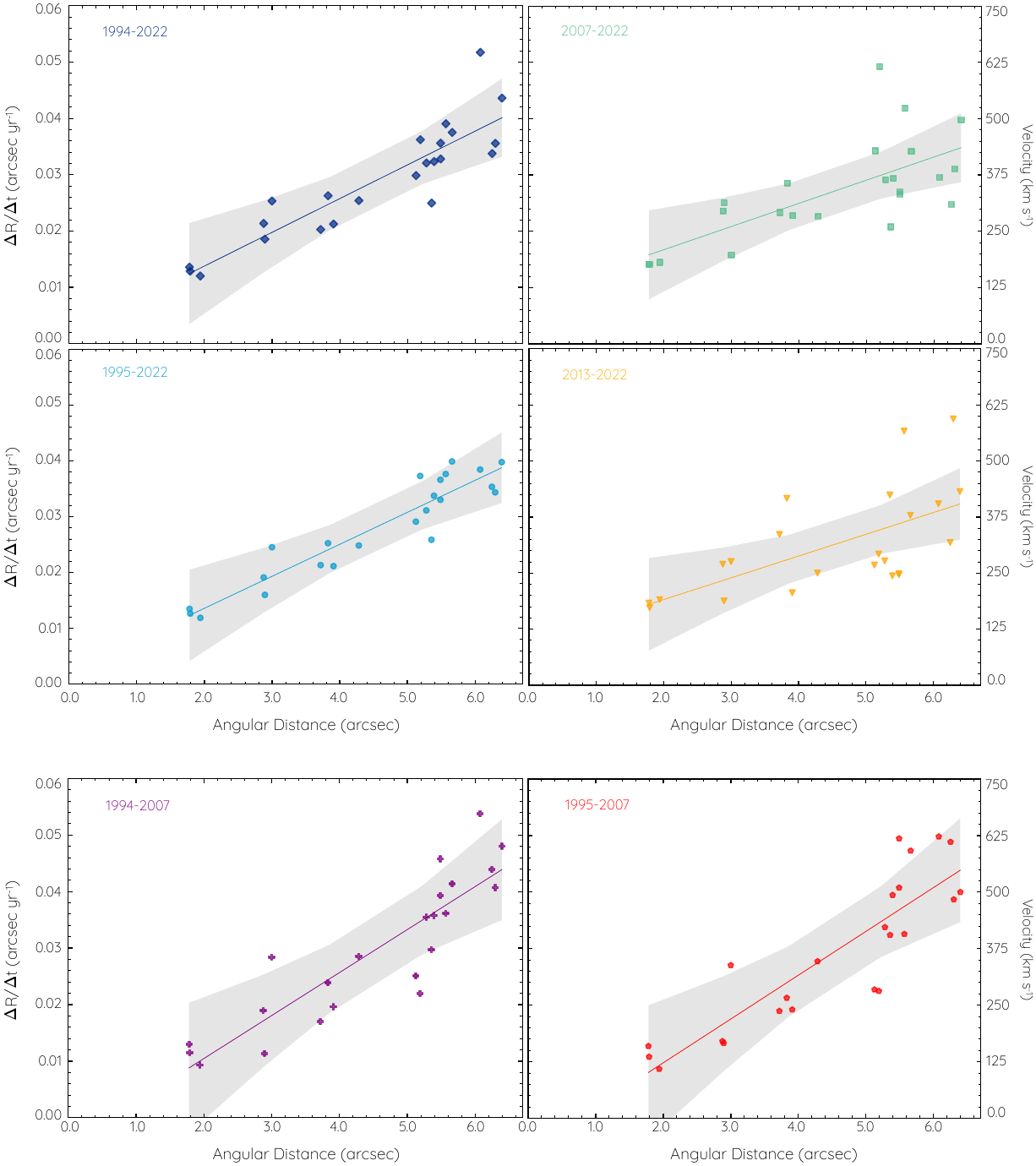}
\caption{Individual fits from bottom panel of Fig.~\ref{fig:radial}. 
The span of both the X and Y axes are respectively identical in all panels to allow a fair comparison of the different fits.}
\label{fig:time}
\end{center}
\end{figure*}

Linear regression fits to the $\Delta R / (\Delta t \times R)$ and $\Delta R / \Delta t$ obtained for each pair of images of T Pyx are presented in Fig.~\ref{fig:acel} and \ref{fig:time}, respectively. In both figures, the grey-shaded areas represent the 95 per cent confidence band. Details of these linear regression fits are listed in Table~\ref{tab:coords2}. The last column of this table lists the Pearson's correlation coefficient ($R-$value) of the measure data. A positive or negative value represents the slope, while values with $|R|>0.5$ correspond to strong correlations.

\begin{table}
\caption{T\,Pyx knots measurements. The labels corresponds to the knots marked on Fig.~\ref{fig:hstmuse} (bottom-left panel).}
\label{tab:coords1}
\centering
\setlength{\tabcolsep}{2.4\tabcolsep}   
\label{tab:coords}
\begin{tabular}{cccc} 
\hline
Knot &   R.A.  &  Dec.    & Radial Distance    \\
 label  &   (J2000.0)  &    (J2000.0)    &  (arcsec)  \\

\hline
1 &  9:04:41.309 & $-$32:22:49.334 & 1.781   \\
2 &  9:04:41.486 & $-$32:22:47.036 & 1.790   \\
3 &  9:04:41.458 & $-$32:22:46.596 & 1.941   \\
4 &  9:04:41.555 & $-$32:22:46.091 & 2.894   \\
5 &  9:04:41.232 & $-$32:22:46.571 & 2.999   \\
6 &  9:04:41.367 & $-$32:22:51.012 & 2.876   \\
7 &  9:04:41.340 & $-$32:22:44.549 & 3.908   \\
8 &  9:04:41.085 & $-$32:22:47.969 & 4.282   \\
9 &  9:04:41.121 & $-$32:22:48.626 & 3.718   \\
10 &  9:04:41.124 & $-$32:22:49.216 & 3.827   \\
11 &  9:04:41.191 & $-$32:22:52.376 & 5.191   \\
12 &  9:04:41.385 & $-$32:22:53.294 & 5.126   \\
13 &  9:04:41.497 & $-$32:22:53.699 & 5.282   \\
14 &  9:04:41.648 & $-$32:22:53.901 & 6.298   \\
15 &  9:04:41.848 & $-$32:22:49.285 & 5.567   \\
16 &  9:04:41.840 & $-$32:22:45.122 & 6.251 \\
17 &  9:04:41.731 & $-$32:22:44.659 & 5.489 \\
18 &  9:04:41.661 & $-$32:22:43.985 & 5.393 \\
19 &  9:04:41.410 & $-$32:22:42.528 & 5.662 \\
20 &  9:04:41.317 & $-$32:22:42.991 & 5.356 \\
21 &  9:04:41.148 & $-$32:22:44.094 & 5.489 \\
22 &  9:04:40.935 & $-$32:22:45.905 & 6.395 \\
23 &  9:04:40.969 & $-$32:22:49.485 & 6.076 \\
\hline
\end{tabular}
\end{table}

\begin{table}
\caption{Linear regression results.}
\label{tab:coords2}
\centering
\setlength{\tabcolsep}{1.2\tabcolsep}   
\label{tab:coords}
\begin{tabular}{lccc} 
\hline
Epoch &   Slope  &  Intercept   & $R$--value   \\
\hline
Figure \ref{fig:acel} \\
\hline
1994-2022 &  $-$0.00018$\pm$0.00013 & 0.00728$\pm$0.00063 & $-$0.28   \\
1995-2022 &  $-$0.00018$\pm$0.00010 & 0.00711$\pm$0.00050 & $-$0.35   \\
2007-2022 &  $-$0.00058$\pm$0.00018 & 0.00891$\pm$0.00086 & $-$0.57   \\
2013-2022 &  $-$0.00072$\pm$0.00020 & 0.00883$\pm$0.00096 & $-$0.61   \\
1994-2007 &  0.00025$\pm$0.00019 & 0.00526$\pm$0.00088 & 0.28   \\
1995-2007 &  0.00018$\pm$0.00015 & 0.00516$\pm$0.00074 & 0.24   \\
\hline
Figure \ref{fig:time} \\
\hline
1994-2022 &  0.00599$\pm$0.00065 & 0.00178$\pm$0.00308 & 0.89   \\
1995-2022 &  0.00572$\pm$0.00046 & 0.00215$\pm$0.00217 & 0.93   \\
2007-2022 &  0.00412$\pm$0.00092 & 0.00847$\pm$0.00436 & 0.69   \\
2013-2022 &  0.00388$\pm$0.00104 & 0.00752$\pm$0.00495 & 0.63   \\
1994-2007 &  0.00760$\pm$0.00087 & $-$0.00473$\pm$0.00413 & 0.88   \\
1995-2007 &  0.00775$\pm$0.00093 & $-$0.00569$\pm$0.00442 & 0.87   \\
All points&  0.00581$\pm$0.00038 & 0.00165$\pm$0.00179 & 0.79   \\
\hline
\end{tabular}
\end{table}


\end{document}